\documentclass[usenatbib]{mn2e}

\usepackage[pdftex]{graphicx}
\usepackage{aas_macros}
\usepackage{times}
\usepackage{amsfonts,amssymb}
\usepackage{thumbpdf}
\usepackage[protrusion,expansion]{microtype}

\newcommand{\amiga}{\texttt{AMIGA}}
\newcommand{\ahf}{\texttt{AHF}}

\newcommand{\mlapm}{\texttt{MLAPM}}
\newcommand{\mhf}{\texttt{MHF}}

\def\ap3m{AP$^3$M}
\def\LCDM{$\Lambda$CDM}

\def\hkpc{$h^{-1}{\ }{\rm kpc}$}
\def\hMpc{$h^{-1}{\ }{\rm Mpc}$}
\def\hMsun{$h^{-1}{\ }{\rm M_{\odot}}$}

\def\nbody{$N$-body}
\def\c15{$c_{\rm 1/5}$}

\def\L30{$\vec{L}_{\rm 30}$}

\def\Rvir{$R_{\rm vir}$}
\def\Mvir{$M_{\rm vir}$}

\def\zform{$z_{\rm form}$}

\def\lesssim{\mathrel{\hbox{\rlap{\hbox{\lower4pt\hbox{$\sim$}}}\hbox{$<$}}}}
\def\gtrsim{\mathrel{\hbox{\rlap{\hbox{\lower4pt\hbox{$\sim$}}}\hbox{$>$}}}}
\long\def\symbolfootnote[#1]#2{\begingroup%
\def\thefootnote{\fnsymbol{footnote}}\footnote[#1]{#2}\endgroup}

\newcommand{\Table}[1]{Table~\ref{#1}}

\newcommand{\Eq}[1]{Eq.~(\ref{#1})}
\newcommand{\Fig}[1]{Figure~\ref{#1}}
\newcommand{\bq}{\begin{equation}}
\newcommand{\eq}{\end{equation}}

\title[Subhalos in WDM Models]{The Dynamics of Subhalos in Warm Dark Matter Models}
\author[Knebe et al.]
       {Alexander Knebe$^1$, Bastian Arnold$^{1,2}$, Chris Power$^3$, \& 
		Brad K. Gibson$^{4,5}$\\
        $^1$Astrophysikalisches Institut Potsdam, An der Sternwarte 16,
		14482 Potsdam, Germany\\
	$^2$Institut f\"ur Astronomie, Universit\"at Wien,
		T\"urkenschanze 17, 1180 Wien, Austria\\
        $^3$Centre for Astrophysics \& Supercomputing, Swinburne
		University, Mail H39, Hawthorn VIC 3122, Australia\\
        $^4$Centre for Astrophysics, University of Central
		Lancashire, Preston PR1 2HE, UK\\
        $^5$School of Physics, University of Sydney, NSW 2006, Australia}

\begin{document}

\newcommand{\dif}{\ensuremath{\mathrm{d}}}
\newcommand{\me}{\ensuremath{\mathrm{e}}}
\newcommand{\mstar}{\ensuremath{M_{*}}}
\newcommand{\AddRef}[1]{\textbf{\it #1}}

\maketitle 
\begin{abstract} 

  We present a comparison of the properties of substructure halos
  (\emph{subhalos}) orbiting within host halos that form in Cold Dark
  Matter (CDM) and Warm Dark Matter (WDM) cosmologies. Our study
  focuses on selected properties of these subhalos, namely their
  anisotropic spatial distribution within the hosts; the existence of
  a ``backsplash'' population; the age-distance relation; the degree
  to which they suffer mass loss; and the distribution of relative
  (infall) velocities with respect to the hosts. We find that the
  number density of subhalos in our WDM model is suppressed relative
  to that in the CDM model, as we would expect.  Interestingly, our
  analysis reveals that backsplash subhalos exist in both the WDM and
  CDM models. Indeed, there are no statistically significant
  differences between the spatial distributions of subhalos in the CDM
  and WDM models. There is evidence that subhalos in the WDM model
  suffer enhanced mass loss relative to their counterparts in the CDM
  model, reflecting their lower central densities. We note also a
  tendency for the (infall) velocities of subhalos in the WDM model to
  be higher than in the CDM model. Nevertheless, we conclude that
  observational tests based on either the spatial distribution or the
  kinematics of the subhalo population are unlikely to help us to
  differentiate between the CDM model and our adopted WDM model.

\end{abstract}

\begin{keywords}
methods: n-body simulations -- galaxies: halos -- galaxies: evolution -- cosmology: theory -- dark matter
\end{keywords}

\section{Introduction}
\label{sec:introduction}

The currently favoured \LCDM\ model of cosmological structure
formation has proven to be extremely successful at describing the
clustering of matter on intermediate to large scales \citep[e.g.,
][]{2005Natur.435..629S, 2007ApJS..170..377S}. In contrast, it has
been argued that the predictions of the \LCDM\ model are at odds with
observations on the scales of galaxies, on the basis of cosmological
\nbody\ simulations. Cold Dark Matter (CDM) halos are predicted to
have ``cuspy'' density profiles with inner logarithmic slopes of
approximately -1.2 \citep[e.g.,][]{2004MNRAS.349.1039N,
  2004ApJ...607..125T, 2005Natur.433..389D, 2005MNRAS.357...82R},
whereas high resolution observations of low surface brightness
galaxies appear to require halos with constant density cores
\citep[e.g., ][]{2007AaA...467..925G, 2007ApJ...659..149M}.
Furthermore, CDM halos are predicted to contain a wealth of
substructure, which we might expect to observe as satellite galaxies
within galactic halos, in sharp contrast to the observed abundance of
satellites around our Galaxy and others
\citep[][]{1999ApJ...522...82K, 1999ApJ...524L..19M}.

Suggested solutions to these problems have included allowing the dark matter
to be collisional (i.e. \emph{self-interacting}) rather than
collisionless \citep{2000PhRvL..84.3760S,2000PhRvD..62d1302B}, 
allowing it to be warm rather than cold 
\citep[][]{2001ApJ...556...93B, 2001ApJ...559..516A,
  2002MNRAS.329..813K}, and introducing non-standard
modifications to an otherwise unperturbed CDM power spectrum
\citep[e.g., ][]{2001astro.ph.11005B, 2003MNRAS.341..617L}. Arguably
the most promising (and least intrusive) modification to the dark
matter paradigm is to allow the dark matter particle to be warm. In
such a case, warm dark matter particles will have a relatively high thermal
velocity dispersion at decoupling and therefore a non-negligible
\emph{free-streaming scale} $\lambda_{\rm fs}$. This modification
results in a change to the primordial matter power spectrum,
corresponding to a damping of density perturbations on scales below
a \emph{filtering scale} $R_f$ (which is related to the
free-streaming scale $\lambda_{\rm fs}$), which in turn is related to
a filtering mass $M_f$ \citep{1986ApJ...304...15B,
  2001ApJ...556...93B, 2001ApJ...559..516A,
  2002MNRAS.329..813K}.\\

Previous studies have revealed that WDM can resolve some of the tension 
between theoretical prediction and observation. In particular, the
abundance of substructure halos (hereafter subhalos) is greatly reduced
in WDM models \citep[][]{2001ApJ...556...93B, 2001ApJ...559..516A, 
2002MNRAS.329..813K} compared to the CDM model. However, the
simulations used in these studies did not have sufficient resolution to 
follow the orbits of subhalos in detail, and so these results are
based on a snapshot of the subhalo population, static in time rather
than a dynamic entity.

Over the last decade cosmological \nbody\ simulations have advanced
and reached a stage where it is possible to study the dynamics of well
resolved subhalos and satellite galaxies and use them as a probe of
cosmology. This field -- dubbed ``near-field cosmology'' \citep[e.g.,
][]{2006Sci...313..311B} -- has prompted numerous studies
\citep[e.g.,][]{Warnick.Knebe.Power.2007, 2007arXiv0709.4027C,
  2007MNRAS.379.1464S, 2007ApJ...662L..71F, 2007ApJ...667..859D,
  2007MNRAS.374...16L, 2007MNRAS.378.1531K, 2006MNRAS.369.1253W,
  2006ApJ...650..550A, 2006MNRAS.368..741K, 2005MNRAS.363..146L,
  2005MNRAS.359.1537R, 2004MNRAS.348..333D, 2004MNRAS.355..819G,
  2004MNRAS.351..410G, 2004ApJ...603....7K, 2004ApJ...609..482K,
  2003ApJ...598...49Z}, but as yet there have been no in-depth
comparisons of subhalos in CDM and WDM models. We address this in the
present study, using high resolution resimulations of a set of
``identical'' galaxy clusters (see next section) forming in CDM and 
WDM models to compare and contrast properties of their subhalo populations.

Rather than seeking to reproduce and verify all of the recent results
for subhalos derived from simulations of the CDM model, we focus on
selected properties of the subhalo population. Namely, we set out to
validate the existence of the so-called ``backsplash'' population
reported in \citet{2005MNRAS.356.1327G} \citep[see also][]{2004AaA...414..445M,
  2004ogci.conf..513M, 2000ApJ...540..113B}, the anisotropic spatial
distribution of satellites \citep[e.g., ][]{2007arXiv0706.1350B,
  2006ApJ...650..550A, 2005ApJ...629..219Z, 2005MNRAS.363..146L,
  2004ApJ...603....7K}, the putative age-distance relation \citep[cf.,
][]{2004MNRAS.355..819G, 2004ogci.conf..513M}, and the degree to which
subhalos suffer mass loss. We also examine the relative velocities of
subhalos with respect to their hosts. This is of particular interest
because it has been argued that the collision velocity of the ``Bullet
Cluster'' \citep{2007ApJ...661L.131M, 2002ApJ...567L..27M} may pose
another challenge to CDM \citep[e.g., ][]{2007MNRAS.380..911S,
  2006MNRAS.370L..38H}. While it might be expected that the large
scale tidal field will be more important for collision velocities in
mergers, it is nevertheless interesting to see whether the precise
nature of the dark matter might play a role.  Having characterised
these properties, we can quantitatively address the question of
whether or not the spatial and kinematic properties of subhalos at the
present day could be used to differentiate between the CDM and WDM
models.

\section{The Numerical Simulations}
\label{sec:simulations}

\subsection{The Raw Data}
\label{sec:raw_data}

Our analysis is based on suite of high-resolution \nbody\
simulations. They were carried out using the publicly available
adaptive mesh refinement code \mlapm\ \citep{2001MNRAS.325..845K}
focusing on the formation and evolution of dark matter galaxy clusters
containing of order one million particles, with mass resolution $1.6
\times 10^8$ \hMsun\ and force resolution $\sim$2\hkpc. We
first created a set of four independent initial conditions at redshift
$z=45$ in a standard \LCDM\ cosmology ($\Omega_0 = 0.3,\Omega_\lambda
= 0.7, \Omega_b = 0.04, h = 0.7, \sigma_8 = 0.9$). $512^{3}$ particles
were placed in a box of side length 64\hMpc\ giving a mass resolution
of $m_p = 1.6 \times 10^{8}$\hMsun.  For each of these initial
conditions we iteratively collapsed eight adjacent particles to one
particle reducing our particle number to 128$^3$ particles. These
lower mass resolution initial conditions were then evolved until
$z=0$. Then, as described by \citet{1997MNRAS.290..411T}, for each
cluster the particles within five times the virial radius were tracked
back to their Lagrangian positions at the initial redshift
($z=45$). Those particles were then regenerated to their original mass
resolution and positions, with the next layer of surrounding large
particles regenerated only to one level (i.e. 8 times the original
mass resolution), and the remaining particles were left 64 times more
massive than the particles resident with the host cluster. This
conservative criterion was selected in order to minimise contamination
of the final high-resolution halos with massive particles.

The three warm dark matter halos were simulated using the same
techniques. In fact, the only difference between the CDM and WDM halos
is the functional form of the primordial power spectrum used as an
input for the initial conditions generator. We follow
\citet{1986ApJ...304...15B} and modify the CDM power spectrum by
multiplying it with a damping function $F^2_{\rm WDM}(k)$, where

\begin{equation} \label{eq:filterfunction}
 F_{\rm WDM}(k) = \exp \left[ -\frac{k R_d}{2} -\frac{(k R_d)^2}{2} \right] \ .
\end{equation}

\noindent
Following \citet{1986ApJ...304...15B}, we parameterise the damping scale $R_d$ 
in terms of the warmon density parameter $\Omega_{\rm wdm}$, its mass 
$m_{\rm wdm}$ and the dimensionless Hubble parameter $h$, 

\begin{equation} \label{eq:warmon_mass}
 R_d \simeq 0.074 \left(\frac{m_{\rm
       wdm}}{keV}\right)^{-4/3}\left(\frac{\Omega_{\rm
       wdm}}{0.3}\right)^{1/3}\left(\frac{h}{0.7}\right)^{5/3}h^{-1} \rm Mpc.
\end{equation}

\noindent We adopt a warmon mass of $m_{\rm wdm}=0.5$ keV, which gives a
damping scale of $R_d = $0.186\hMpc. The filtering scale $R_f$ can be obtained 
by determining the wavenumber of the mode at which the amplitude of the linear
density fluctuation is suppressed by a factor of two, and then computing half
the comoving wavelength. It is straightforward to evaluate this from the WDM
and CDM power spectra. For our choice of warmon mass and cosmological
parameters, we find that this wavenumber corresponds to $k=2.553 \,h 
\,{\rm Mpc}^{-1}$ and therefore the filtering scale
(mass) is $R_f=1.24 \,h^{-1} \rm Mpc$ ($M_f=6.653 \times 10^{11} h^{-1} 
{\rm M}_{\odot}$).

Note that our choice of warmon mass is lower than recent lower limits derived 
from combined analysis of observed properties of the matter power spectrum as 
inferred from the Sloan Digital Sky Survey Lyman-$\alpha$ flux power spectrum, 
cosmic microwave background data and the galaxy power spectrum, which vary 
between, e.g., $m_{\rm wdm} \geqslant 3$ keV \citep{2006PhRvD..73f3513A} and 
$m_{\rm wdm} \geqslant 10$ keV \citep{2006PhRvL..97g1301V}. However, it is
consistent with published estimates of the $\sim 0.5$ keV warmon mass that 
would resolve the ``overabundance'' of dark matter substructure in galactic 
halos
\citep[e.g.][]{1999ApJ...524L..19M,2001ApJ...561...35D,2006MNRAS.368.1073G}.
By focusing on the lowest warmon mass that could be considered consistent with
observational data, we can explore subhalo dynamics in a plausible model, yet
one in which the effects of the warmon should be more pronounced and therefore 
easier to identify when comparing and contrasting with the CDM model.

We note that the three CDM halos CDM1, CDM2, and CDM3 have appeared previously 
in both \citet{2006MNRAS.369.1253W} and \citet{Warnick.Knebe.Power.2007}, in 
which they corresponded to the ``C3'', ``C7'' and ``C8'' systems.

\subsection{Discreteness Effects}
\label{sec:discreteness_effects}
It has been argued in the recent study of \citet{2007MNRAS.380...93W}
that WDM halos below a given fraction of the filtering mass
$M_f$ are spurious, arising from the unphysical fragmentation of
filaments. They provided the following expression (based upon
simulations of the Hot Dark Matter model)

\begin{equation}
\label{eq:wang}
{\rm M_{\rm lim}} \simeq 10.1 \overline{\rho}\,d\,k_{\rm peak}^{-2} \ .
\end{equation}

\noindent $\overline{\rho}$ is the mean density, $d$ is the mean
interparticle separation, and $k_{\rm peak}$ is the wavenumber at
which $\Delta^2(k) = k^3 P(k)$ reaches its maximum. 

We have taken care to compute $\rm M_{\rm lim}$, noting that our
simulations use boxes of side $64 h^{-1} \rm Mpc$, an effective number
of particles $512^3$, a density parameter of $\Omega_0$=0.3 and a peak
wavenumber of $k_{\rm peak}$=1.78 $h$ Mpc$^{-1}$\symbolfootnote[1]{The peak
  value $k_{\rm peak}$=1.78 $h$ Mpc$^{-1}$ has been determined
  numerically from the (tabulated) warm dark matter power spectrum
  used in this study based upon a CDM power spectrum calculated with
  the publically available \texttt{CMBFAST} code
  \citep{1996ApJ...469..437S} and modified according to
  \Eq{eq:filterfunction}.}.  These numbers give $d$=0.125 $h^{-1} \rm
Mpc$, $\overline{\rho}$=8.3265 $h^2\,{\rm M_{\odot}}\,{\rm Mpc}^{-3}$
and so $\rm M_{\rm lim}$=$3.317828\times 10^{10} h^{-1} \,\rm
M_{\odot}$. Our particle mass is $m_p=1.626269\times 10^8 h^{-1} \,\rm
M_{\odot}$, and so we find that $\rm M_{\rm lim}$ is equivalent to 204
particles. Therefore, this corresponds to the mass cut applied in the
following analysis. 

We have also applied a cut of 2$\rm M_{\rm lim}$ to our data and checked 
our results to ensure that they are unaffected by spurious halos. This
reveals that our results remain unchanged and therefore are stable. We do not
find any systematics biases in our data if we employ mass cuts of $\rm M_{\rm
  lim}$ or 2$\rm M_{\rm lim}$. We have looked for trends in our data that we
would expect to be present if they were affected by spurious haloes (such as
the distribution of concentrations) but we find no obvious signatures.
Therefore we conclude that our results are robust and unaffected by particle
discreteness. 

\subsection{The Halos}
\label{sec:host_halos}

Both the halos and their subhalos are identified using
\ahf\footnote{{\small \textbf{A}MIGA}'s-{\small
    \textbf{H}}alo-{\small\textbf{F}}inder; \ahf\ can be downloaded
  from \texttt{http://www.aip.de/People/aknebe/AMIGA}. \amiga\ is the
  successor to \mlapm.}, a modification of the
\mhf\footnote{\mlapm's-\texttt{H}alo-\texttt{F}inder} algorithm 
presented in \cite{2004MNRAS.351..399G}, which has been parallelised
using the MPI (Message Passing Interface) libraries. \ahf~utilises the adaptive
grid hierarchy of \mlapm\ to locate (sub)halos as peaks in an
adaptively smoothed density field. Local potential minima are computed
for each peak and the set of particles that are gravitationally bound
to the peak are returned. If the peak contains in excess of 20
particles, then it is considered a (sub)halo and it is retained for
further analysis.

For each (sub)halo we calculate a suite of canonical properties
from particles within the virial/truncation radius. We define the
virial radius $R_{\rm vir}$ as the point at which the density profile
(measured in terms of the cosmological background density $\rho_b$)
drops below the virial overdensity $\Delta_{\rm vir}$, i.e. $M(<R_{\rm
  vir})/(4\pi R_{\rm vir}^3/3) = \Delta_{\rm vir} \rho_{\rm b}$. Here 
$\rho_{\rm b}$ is the mean density of the background (Universe). Following
convention, we assume the cosmology- and redshift-dependent definition
of $\Delta_{\rm vir}$; for a distinct (i.e. host) halo in a \LCDM\
cosmology with the cosmological parameters that we have adopted,
$\Delta_{\rm vir}=340$ at $z=0$. This prescription is not appropriate
for subhalos in the dense environs of their host halo, where the
local density exceeds $\Delta_{\rm vir} \rho_{\rm b}$, and so the density
profile will show a characteristic upturn at a radius $R \lesssim
R_{\rm vir}$. In this case we use the radius at which the density
profile shows this upturn to define the truncation radius for the
subhalo. Further details of this approach (and especially the ``halo
tracking'' used to obtain the temporal evolution of subhalos) can be
found in \cite{2004MNRAS.351..399G}.

In \Table{tab:hosts} we summarise some of the properties of the host
halos along with particulars of their respective subhalo populations.

\begin{table*}
  \caption{Summary of the host halos properties and their subhalo populations.
    The age is given in Gyrs, \Rvir\ is measured in \hkpc, masses in 
    $10^{14}$\hMsun\, and the velocity dispersion $\sigma_v$ in
    km/sec. We follow \citet{1993MNRAS.262..627L} and define the formation redshift
    of our host halos as the redshift at which the halo's most massive progenitor
    first contains in excess of half its present day mass. The concentration
    $c_{1/5}=R_{\rm vir}/R_{1/5}$ is defined via the radius 
    $R_{1/5}$ that encompasses $1/5^{\rm th}$ of the virial mass. Shape is quantified by
    the triaxiality parameter $T=(a^2-b^2)/(a^2-c^2)$ and the
    eigenvalues of the inertia tensor $a>b>c$.  $N_{\rm sat}^{X}$
    measures the number of a certain subset $X$ ($\rm int$=interior, 
    $\rm inf$=infalling, $\rm back$=backsplash) of subhalos 
    while $N_{\rm sat}^{< 2.5 R_{\rm vir}}$ gives
    the total number of subhalos within 2.5\Rvir; note that these
    numbers only reflect subhaloes in excess of 200 particles.}
  \label{tab:hosts}
  \begin{tabular}{@{}ccccccccccccccc}
    \hline
    model                                &
    \zform\                              &
    age                                  &
    \Rvir\                               &
    \Mvir\                               &
    $\sigma_v$                           &
    $c_{1/5}$                             &
    $T$                                  &
    $b/a$                                &
    $c/a$                                &
    max\{$M_{\rm sat}$\}                  &
    $N_{\rm sat}^{<2.5R_{\rm vir}}$         &
    $N_{\rm sat}^{\rm int}$               &
    $N_{\rm sat}^{\rm inf}$              &
    $N_{\rm sat}^{\rm back}$  \\
    \hline
    
    CDM1 & 0.805 & 6.9 & 973  & 1.1 &  833 & 6.57 &   0.952 &   0.749 &   0.212 & 0.12 & 54 & 26  & 13  & 15  \\
    CDM2 & 0.443 & 4.6 & 1347 & 2.9 & 1185 & 5.91 &   0.836 &   0.597 &   0.467 & 0.18 & 182 & 91 & 53 & 38 \\
    CDM3 & 0.237 & 2.8 & 1379 & 3.1 & 1092 & 5.84 &   0.867 &   0.818 &   0.749 & 0.49 & 159 & 126 & 31 & 2  \\
\\
    WDM1 & 0.871 & 7.1 & 967  & 1.1 &  783 & 5.98 &   0.958 &   0.773 &   0.206 & 0.09 & 17  & 6  & 7  & 4  \\
    WDM2 & 0.643 & 5.9 & 1340 & 2.8 & 1093 & 6.09 &   0.887 &   0.705 &   0.423 & 0.45 & 77 & 52  & 16  & 9  \\
    WDM3 & 0.284 & 3.2 & 1352 & 3.0 & 1119 & 3.69 &   0.837 &   0.693 &   0.576 & 0.36 & 68 & 50 & 12  & 6  \\
    \hline
  \end{tabular}
  
\end{table*}

\section{Analysis of the Subhalo Population}
\label{sec:analysis}

\subsection{The ``Backsplash'' Population}
\label{sec:backsplash}

\begin{figure}
  \includegraphics[width=84mm]{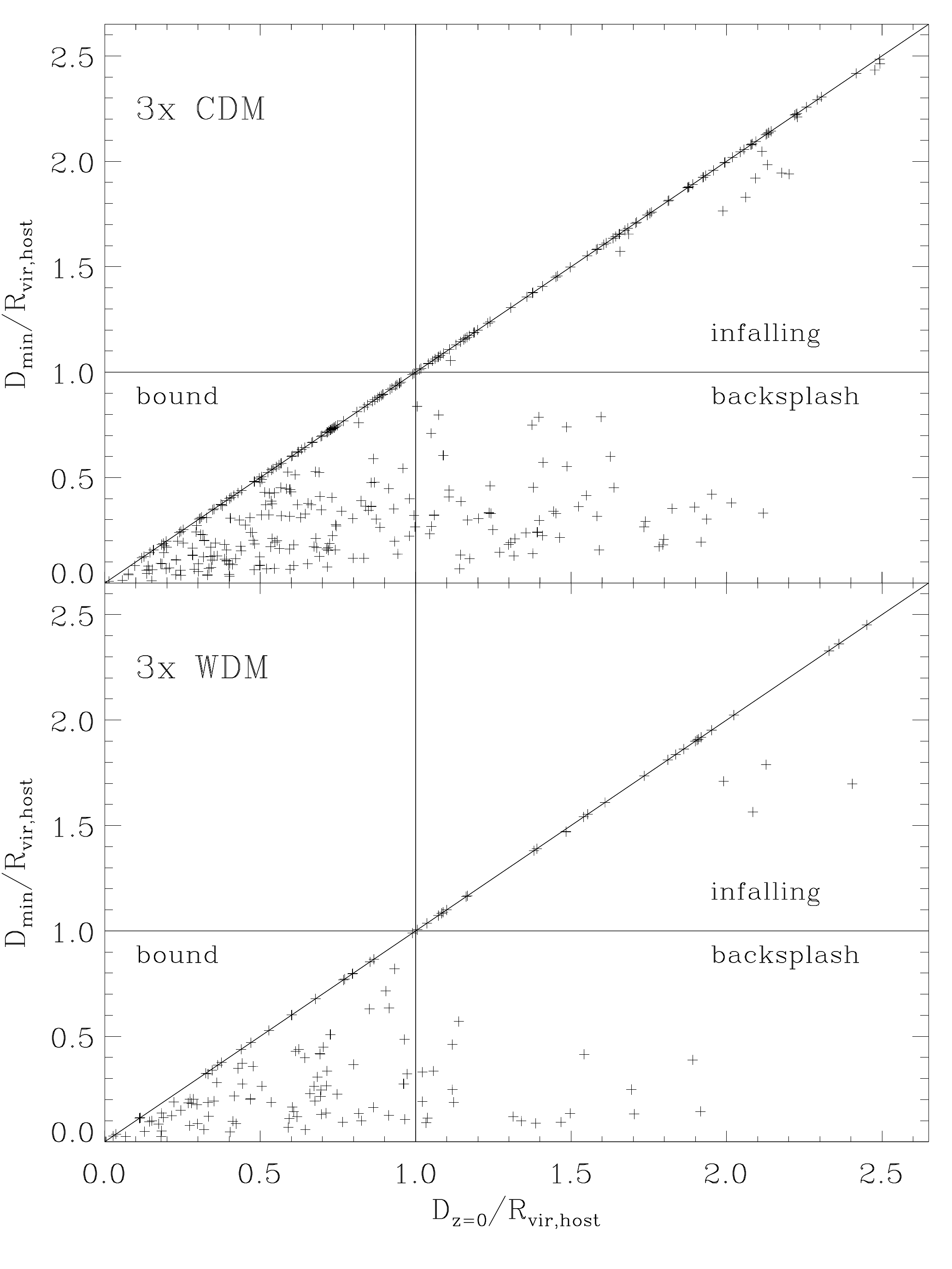}
  \caption{Minimum distance as a function of present-day distance.}
  \label{fig:Dmin}
\end{figure}

It has been noted that a significant population of halos on the
outskirts of present-day galaxy- and cluster-mass host halos once
resided within the virial radii of these hosts at earlier times
\citep{Warnick.Knebe.Power.2007, 2005MNRAS.356.1327G,
  2004AaA...414..445M, 2004ogci.conf..513M, 2000ApJ...540..113B}.
These results are based on simulations of the CDM model, but it is
interesting to ask whether such a ``backsplash'' population exists in
the WDM model. We expect there to be fewer satellites in WDM models
and these satellites will tend to have lower concentrations
\citep{2002MNRAS.329..813K, 2001ApJ...559..516A, 2001ApJ...556...93B},
which, when combined, should affect the numbers of ``backsplash'' halos.
The lower the concentration of a (sub)halo, the greater the likelihood
that it will be tidally disrupted within the host. Therefore, we might
expect the numbers of backsplash halos to be suppressed in WDM models.

In \Fig{fig:Dmin} we plot the minimum halocentric distance reached by
a (sub)halo against its present day halocentric radius. Note that we
combine data for all three halos in the CDM and WDM models
respectively. As expected, backsplash halos are present in both the
CDM and WDM models, although the numbers are reduced in the WDM model.
\Table{tab:hosts} reveals that the youngest system (i.e. host \#3) has
the smallest fraction of backsplash halos; this reflects the fact that
this system has experienced a recent triple merger
\citep[cf.][]{Warnick.Knebe.Power.2007,
  2006MNRAS.369.1253W}. Interestingly, we find again
\citep[cf. ][]{2005MNRAS.356.1327G} that the number of infalling halos
is of the same order as the number of backsplash halos in both the CDM
and WDM models.

\subsection{Mass Spectra}
\label{sec:mass_spectra}
\begin{figure}
  \includegraphics[width=84mm]{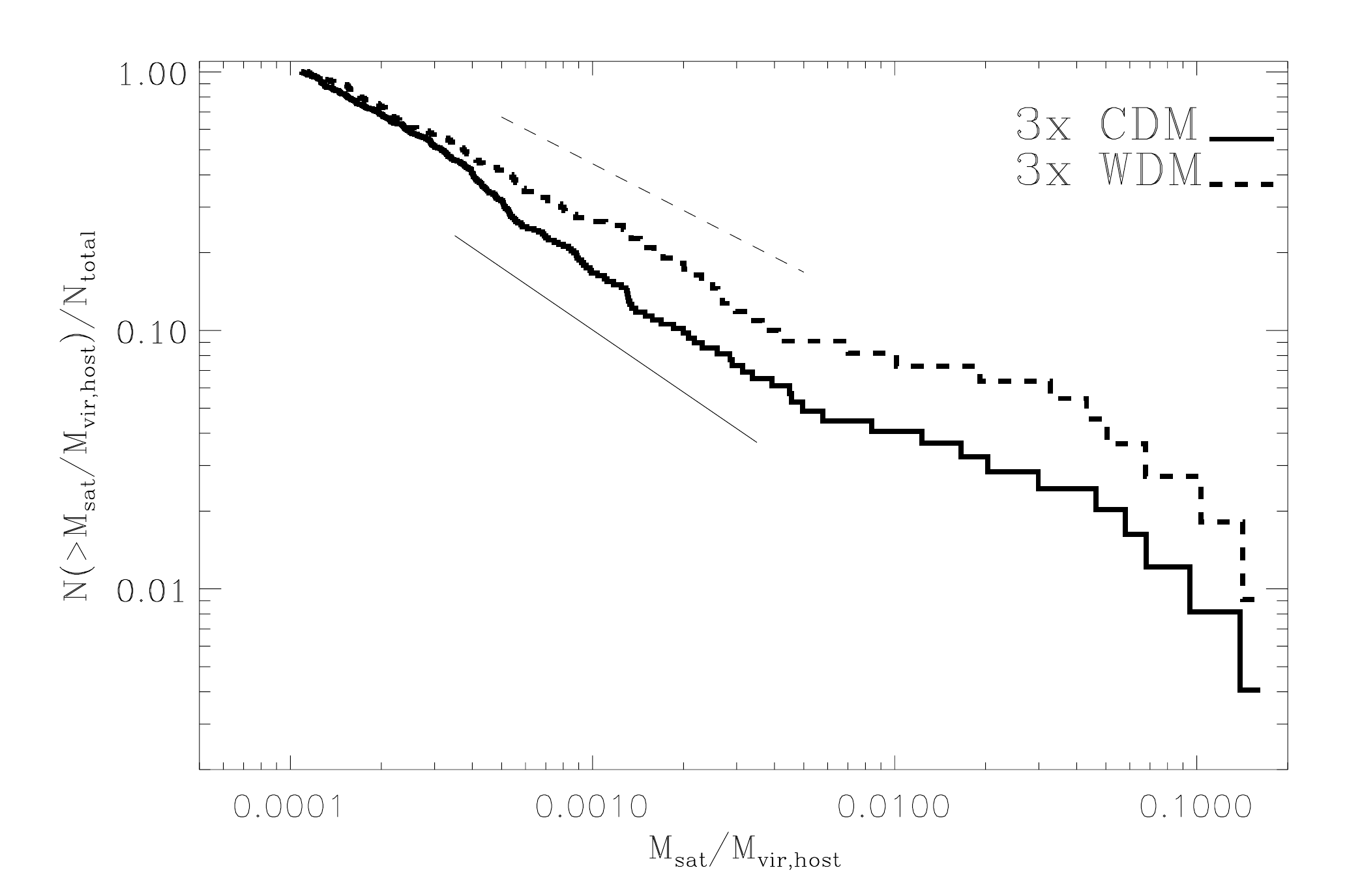}
  \caption{The cumulative distribution of all interior subhalos normalised 
    to the total number of satellites. The short thin lines represent
    curves with the logarithmic slopes of $\alpha=-0.8$ (solid) and 
    $\alpha=-0.6$ (dashed), respectively.}
  \label{fig:CumMass_bound_logX}
\end{figure}

\begin{figure}
  \includegraphics[width=84mm]{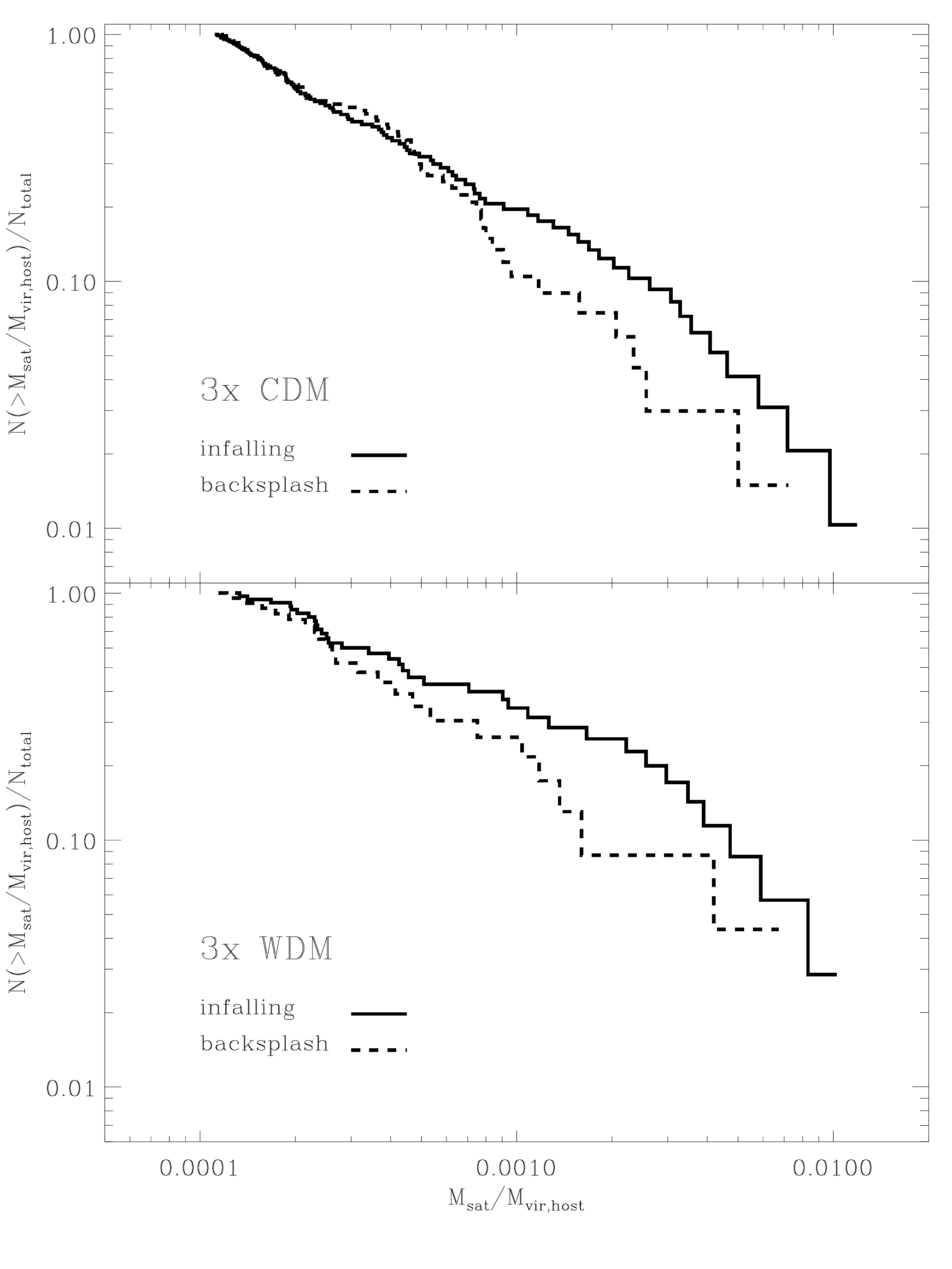}
  \caption{As in \Fig{fig:CumMass_bound_logX} but for the infalling
    and backsplash populations respectively.}
  \label{fig:CumMass_infback_logX}
\end{figure}

The mass spectrum of satellite galaxies in WDM and CDM has been studied 
previously \citep{2002MNRAS.329..813K,
  2001ApJ...559..516A, 2001ApJ...556...93B}, but we consider it 
here briefly for completeness. In \Fig{fig:CumMass_bound_logX} we show the 
cumulative mass functions of subhalos normalised to the total 
number of subhalos, where we normalise the subhalo mass by the host halo
mass. It is readily apparent that the abundance of low-mass halos is 
suppressed in WDM cosmologies. We fit a power-law to this mass function,

\begin{equation}
  \frac{N(>M)}{N_{\rm total, interior}} \propto \left( \frac{M_{\rm sat}}{M_{\rm vir, host}} \right)^\alpha,
\end{equation}

\noindent and obtain logarithmic slopes of $\alpha = -0.8$ for CDM and
$\alpha = -0.6$ for WDM. This is consistent with the findings of
previous studies for CDM subhalos \citep[e.g.,
][]{2007ApJ...659.1082S, 2005MNRAS.359.1537R, 2004MNRAS.351..399G,
  2004MNRAS.348..333D, 2004MNRAS.355..819G}.

We also calculate the mass functions of the infalling and backsplash 
halo populations, which we show in \Fig{fig:CumMass_infback_logX}. We 
observe a general trend that backsplash halos contain fewer high-mass 
objects in comparison to the infalling halos. This reflects the 
importance of tidally induced mass loss for backsplash halos, which we 
quantify in the next section \citep[cf. also ][]{Warnick.Knebe.Power.2007}.

\subsection{Mass Loss}
\label{sec:mass_loss}
\begin{figure}
  \includegraphics[width=84mm]{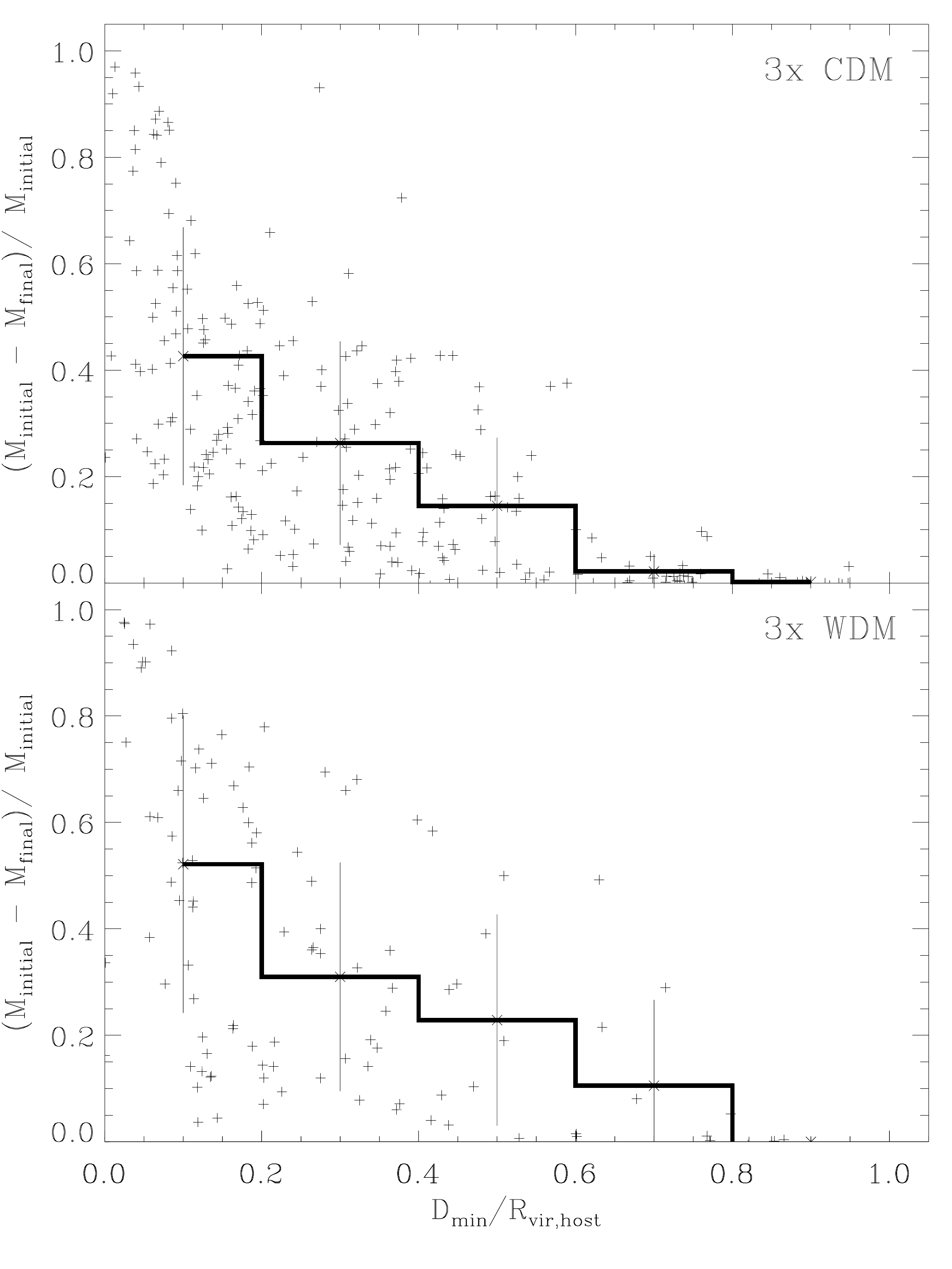}
  \caption{The total (fractional) mass loss of interior subhalos as a 
	function of minimum distance to the host. The error bars
        represent the standard deviation.}
  \label{fig:MassLoss_bound_stack}
\end{figure}

\begin{figure}
  \includegraphics[width=84mm]{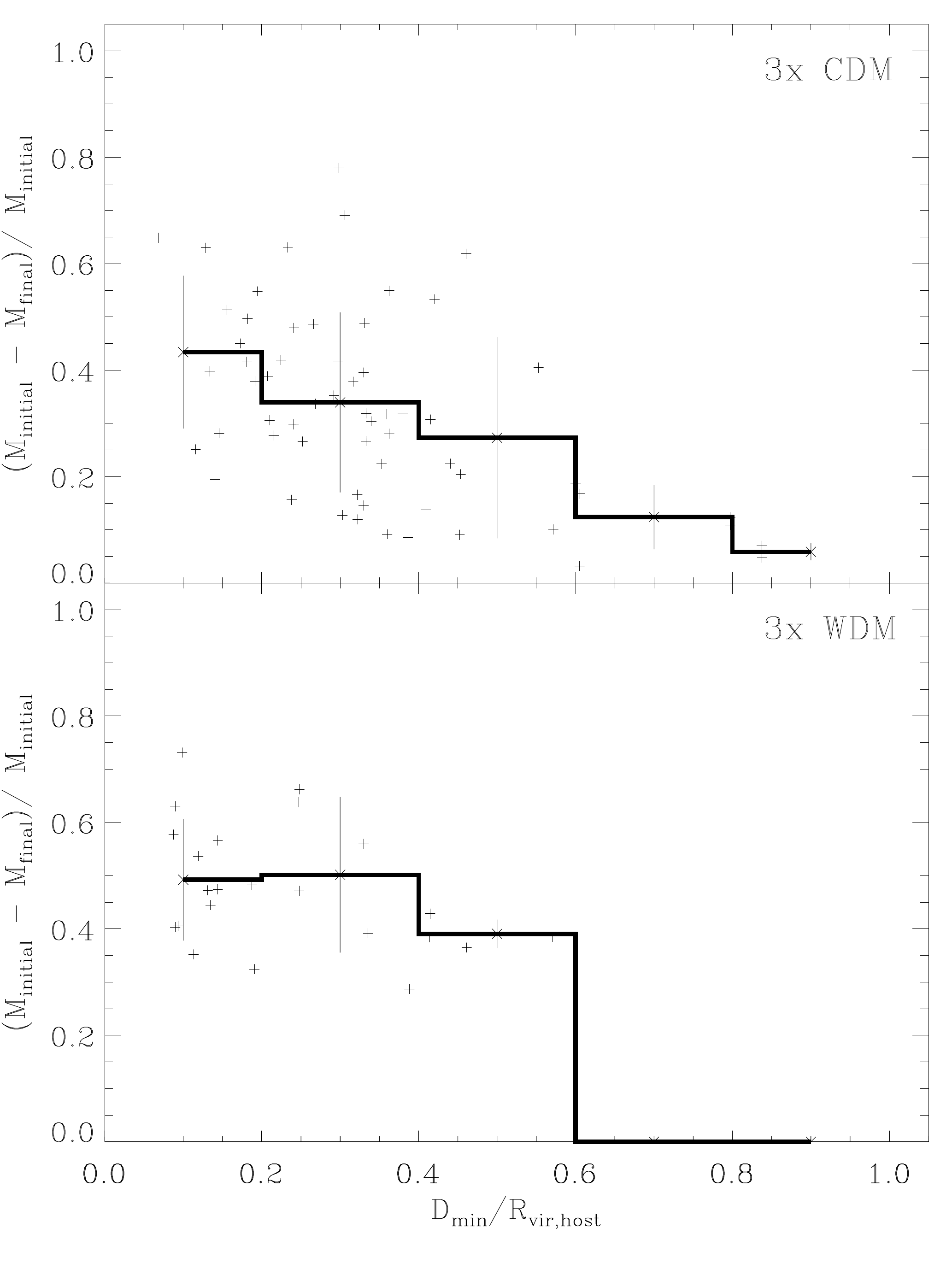}
  \caption{Same as \Fig{fig:MassLoss_bound_stack} but for the backsplash 
	population.}
  \label{fig:MassLoss_backsplash_stack}
\end{figure}

There is a general consensus that subhalos in WDM models are less
concentrated than their counterparts in CDM models
\citep{2002MNRAS.329..813K, 2001ApJ...556...93B,2001ApJ...559..516A}
and therefore more susceptible to tidal destruction while orbiting
within the dense environs of their host halo. We verify this in
\Fig{fig:MassLoss_bound_stack} and \Fig{fig:MassLoss_backsplash_stack}
where we plot the total (fractional) mass loss as a function of
distance to the host for both the interior and the backsplash
population. The mass loss is measured over the time period from infall
onto the host (i.e. the first time a satellite crosses the virial
radius of the host on an inward trajectory) until the present-day.

In both dark matter models the average mass loss (presented as histograms 
in \Fig{fig:MassLoss_bound_stack}) is a monotonic decreasing 
function of minimum distance. However, in the WDM model this function is 
pointwise greater than the corresponding curve in CDM, which would be
expected if mass loss is enhanced as a result of the lower concentrations 
of subhalos in the WDM model.

Surprisingly, the relation between mass loss and minimum distance is
not as steep for backsplash halos as for interior subhalos
(\Fig{fig:MassLoss_backsplash_stack}). Nevertheless, (sub)halos
plunging deeper into the potential well of the host experience greater 
mass loss -- as expected and confirmed for the interior population 
\citep[e.g.,][]{2007MNRAS.379.1464S,2007ApJ...667..859D,2004ApJ...609..482K, 
2004MNRAS.355..819G}. On average, backsplash halos in the WDM model
suffer greater mass loss than their CDM counterparts.

\subsection{Spatial Anisotropy} 
\label{sec:spatial_anisotropy} 
\begin{figure}
  \includegraphics[width=84mm]{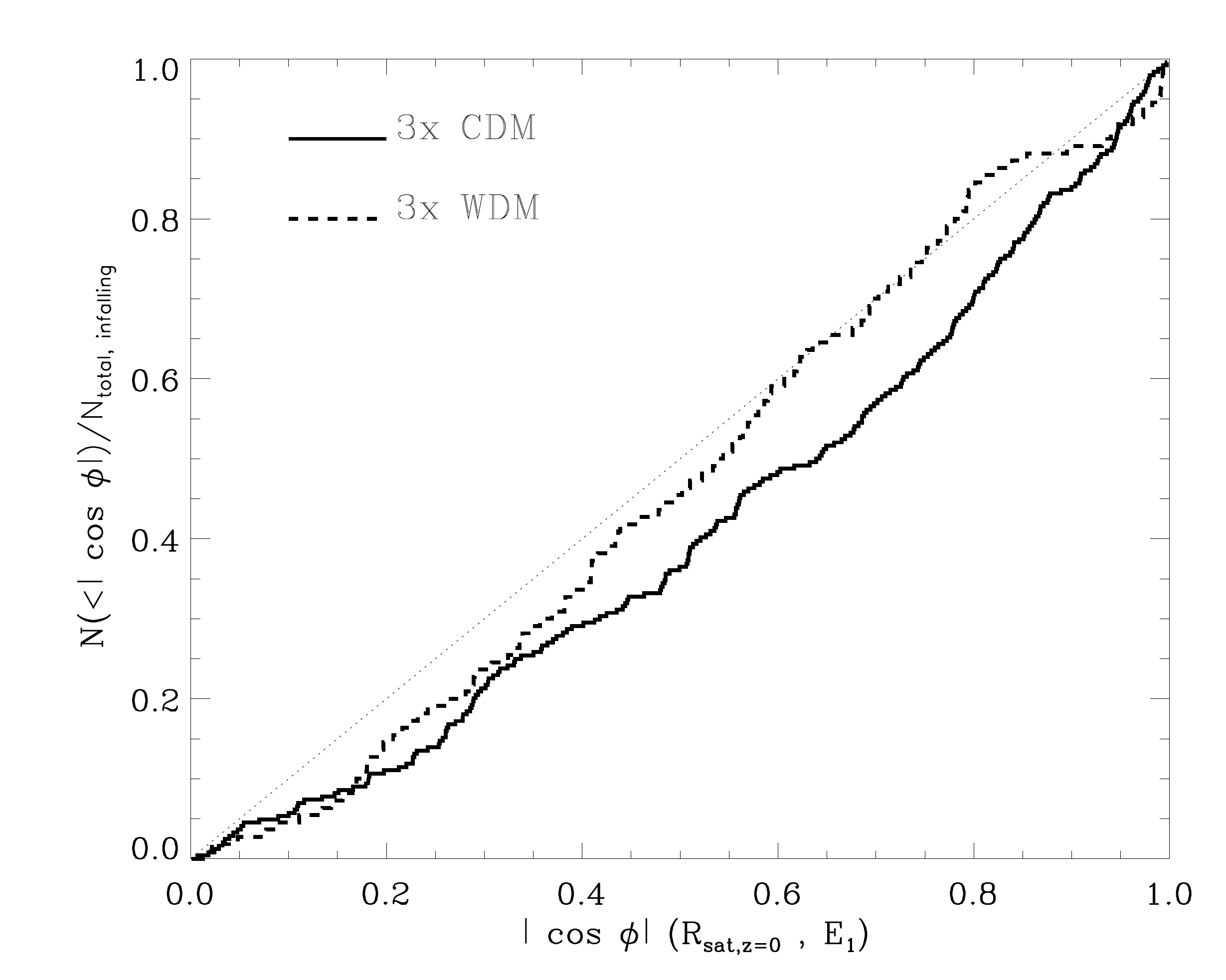}
  \caption{Cumulative fraction of satellites with the absolute value
    of the cosine of the zenith angle $< |\cos \phi|$. The zenith
    angle, $ 0 < \phi < \pi$, is defined as the angle from the major
    axis of the dark matter distribution of the host. The dotted line
    corresponds to an isotropic distribution or the ``uniform
    continuous distribution function''.}
  \label{fig:Anisotropy_z0_bound}
\end{figure}

\begin{figure}
 \begin{minipage}{0.4\textwidth}
  \includegraphics[width=80mm]{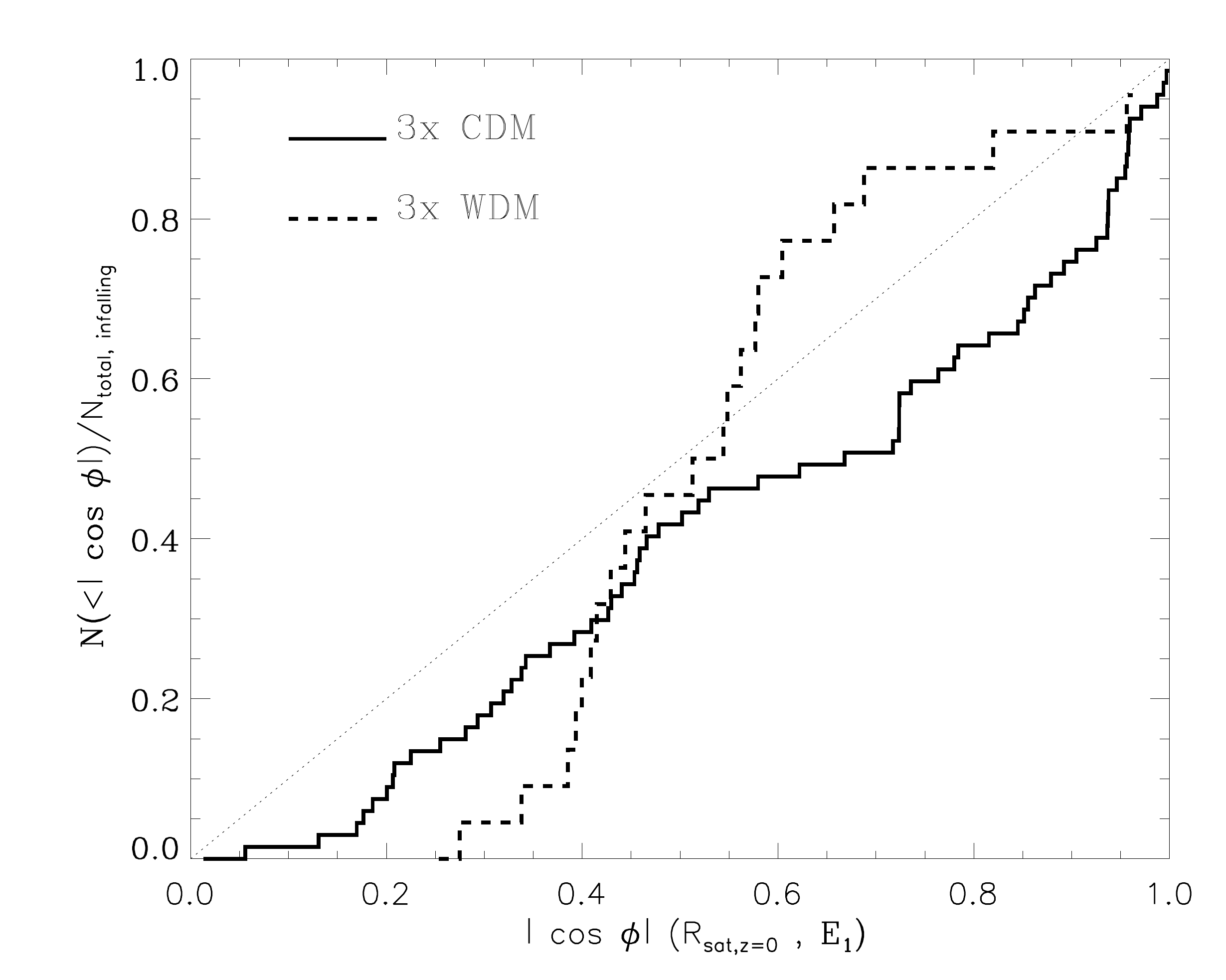}
 \end{minipage} 

 \begin{minipage}{0.4\textwidth}
  \includegraphics[width=80mm]{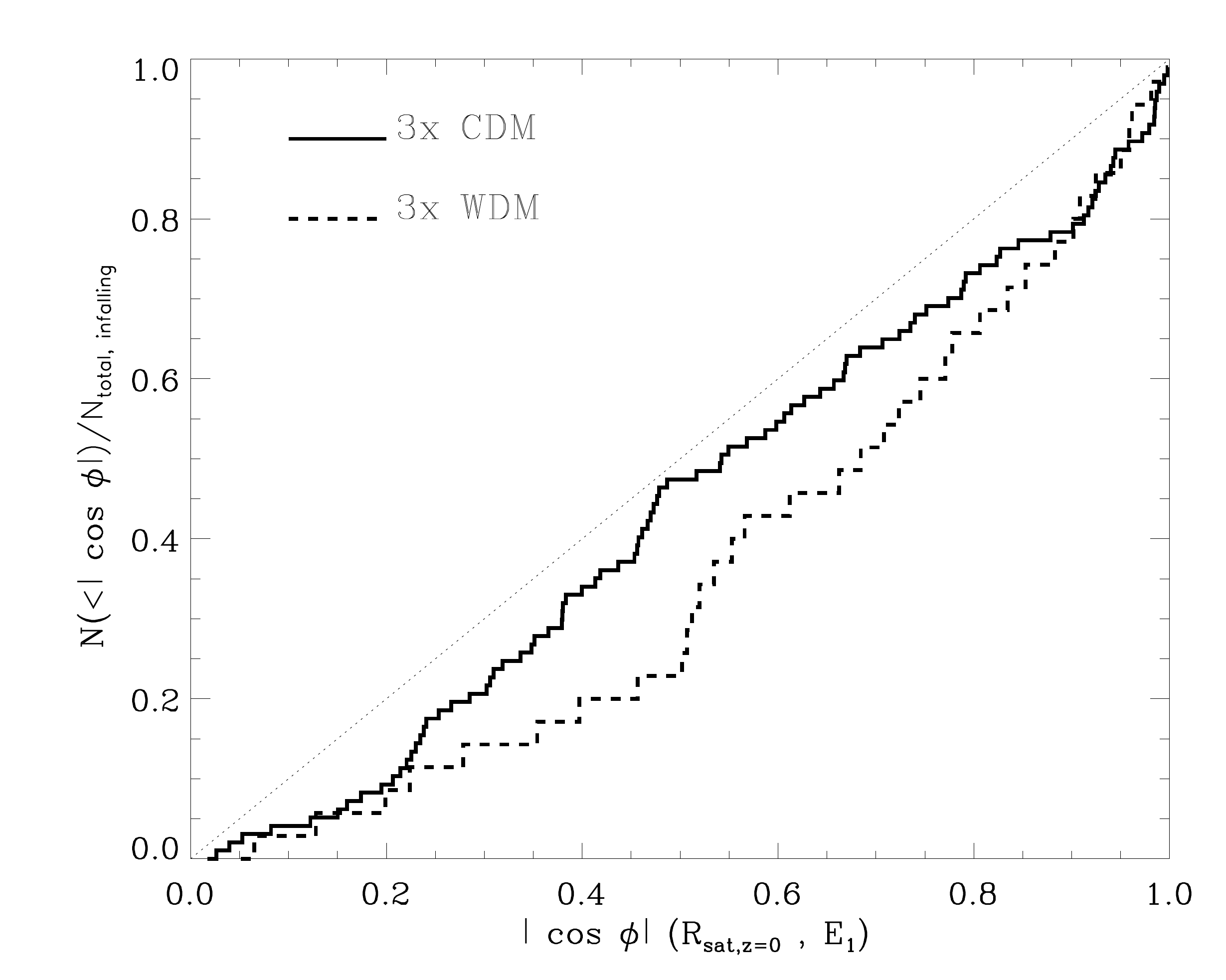}
 \end{minipage}
  \caption{Same as \Fig{fig:Anisotropy_z0_bound} but for the backsplash 
	(upper panel) and infalling (lower panel) population, respectively.}
  \label{fig:Anisotropy_z0_backsplash_infalling}
\end{figure}

There is good reason to believe that the spatial distribution of
subhalos (and the subset corresponding to satellite galaxies) in 
both cluster sized systems and galactic halos is anisotropic in the CDM
model \citep[e.g.,][]{2007ApJ...662L..71F, 2007MNRAS.378.1531K,
  2007MNRAS.374...16L, 2007arXiv0704.3441A, 2006ApJ...650..550A,
  2005ApJ...629..219Z, 2005MNRAS.363..146L, 2004ApJ...603....7K}, and
so it is interesting to ask whether the same can be said of the WDM
model.

For each of our host halos, we compute the cumulative fraction of
subhalos which have cosine of the angle

\begin{equation}
 \cos{\Phi} = {\bf R_{\rm sat, z=0}} \cdot {\bf E_1};
\end{equation}

\noindent this measures the position of a subhalo relative to the host
${\bf R_{\rm sat, z=0}}$ and the host's major axis ${\bf E_1}$. The
host's major axis is identified using the eigenvalues and eigenvectors
of its moment of inertia tensor, where the eigenvector corresponding
to the smallest eigenvalue defines the major axis.

The resulting distributions of $\cos{\Phi}$ are shown in
\Fig{fig:Anisotropy_z0_bound} for subhalos within the host's virial
radius, which confirms that the spatial anisotropy is present in the
WDM model although not as pronounced as for the CDM case. Although we
do not show the result here as it (probably) lies below the
credibility level of the WDM simulation \citep{2007MNRAS.380...93W},
we note that very low mass systems (i.e. $M_{\rm sat}<10^{-4}M_{\rm
  host}$) correlate more strongly with the major axis than the
remainder of the satellites -- for both dark matter models. This is
consistent with the expectation that (especially low mass) objects are
primarily channelled along the filaments feeding the cluster
\citep[cf. also ][]{2004ApJ...603....7K}. The thin dashed line
represents an isotropic distribution or the ``uniform continuous
distribution function'' (UCDF).

We have computed the same distribution for both the backsplash and
infalling satellites in \Fig{fig:Anisotropy_z0_backsplash_infalling}.
Interestingly we find that the spatial anisotropy is even stronger for
the backsplash population than for the interior objects even though it is
skewed towards $\cos\Phi\approx0.55$ for the WDM model and hence no
perfect alignment with $E_1$ anymore (but nevertheless an anisotropic
distribution). One possible explanation for this could be that
backsplash halos tend to be on radial orbits that are either plunging
through the host or grazing the ``virial surface''. If this is the
case, we might expect to observe a tendency for the infalling
population to align with the major axis as verified in the lower panel
of \Fig{fig:Anisotropy_z0_backsplash_infalling}. In addition we note
that the infalling WDM satellites show a more pronounced tendency to
be aligned with the major axis of the respective host. This goes along
with the expectation for (sub)haloes to be concentrated in the
filaments and channelled along them into the cluster \citep[e.g.,
][]{2004ApJ...603....7K}.

\subsection{Age-Distance Relation}
\label{sec:age_distance}

\begin{figure}
  \includegraphics[width=84mm]{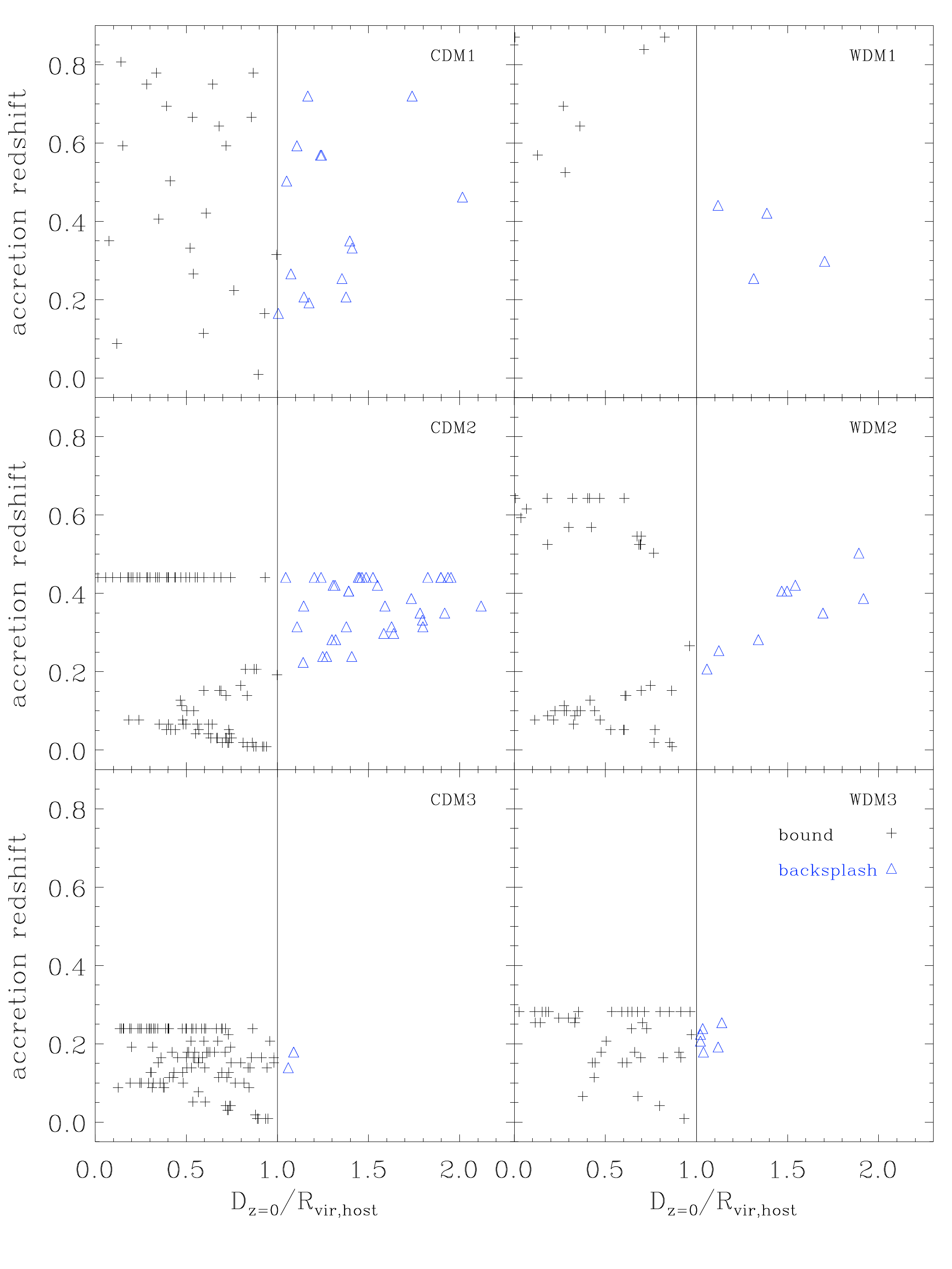}
  \caption{Age-Distance Relation. We define the ``age'' of a subhalo
    as the period of time that has elapsed since it first crossed the
    virial radius of its host on an inward trajectory. The redshift at
    which this occurs is the ``accretion redshift''.  $D_{z=0}/R_{\rm
      vir, host}$ measures the subhalo's current distance to with
    respects to the centre of the host.  For clarity, the triangles
    represent backsplash halos while black crosses denote ``interior''
    subhalos.}
  \label{fig:ZredDist_backnbound}
\end{figure}

Is there is a correlation between the infall time of a satellite and
its present-day halocentric radius? Previous studies using
cosmological simulations have argued that ``older'' subhalos tend to
lie closer to the centre of the host \citep[][]{2004MNRAS.353..639W,
  2004MNRAS.355..819G}, but there are also claims to the contrary
\citep[][]{2004ogci.conf..513M}. Here the age of a subhalo corresponds
to the period of time that has elapsed since it was first accreted by
the host (accretion redshift), entering the virial radius on an inward
trajectory.

We investigate whether such a correlation between a subhalo's age or
\emph{accretion redshift} and halocentric radius exists in our data in
\Fig{fig:ZredDist_backnbound}. Crosses represent ``interior'' subhalos
within the virial radius at the present day while triangles represent
backsplash halos.  Because we output our snapshots at discrete
intervals, the times at which subhalos are accreted appear
discrete. It is possible to correct for this discreteness by
interpolating the growth of the virial radius and the positions of
subhalos between snapshots, but we output snapshots sufficiently
frequently to make the uncertainty introduced by discreteness
negligible.

\Fig{fig:ZredDist_backnbound} reveals that the correlation between
radius and age is not a straightforward one. The CDM1/WDM1 system is hard to
interpret because there is no strong trend for subhalos within the virial
radius. However, there are interesting trends in the CDM2/WDM2 and
CDM3/WDM3 that suggest that there may be distinct populations following
distinct age-distance relations. The most recently accreted subhalos, with 
accretion redshifts $z \lesssim 0.1$, show the expected trend for
accretion redshift to increase with decreasing redshift. However,
subhalos accreted at $z \gtrsim 0.1$ appear to follow an inverse
relation, tending to have higher accretion redshifts for larger
halocentric radii, and this trend continues beyond the virial radius 
into the backsplash population. Finally, we note that the ``oldest''
subhalos do not appear to follow any trend, instead forming a hard
upper edge in each panel. However, this edge is an artifact of our
method for tracking subhalos \citep[cf. ][]{2004MNRAS.351..399G} and 
corresponds to the formation redshift, $z_{\rm form}$, of the
respective host halo. This explains why there is a systematic shift to 
higher redshifts from the lower plot to the upper one -- our halo tracker
starts following subhalos at $z_{\rm form}$ and so it cannot ``see'' 
(sub)halos prior to $z_{\rm form}$. Therefore all subhalos that 
resided within the host at this initial time appear as infalling ones.

This figure also reveals that no backsplash halos have been accreted more 
recently than $z\approx 0.15$, which corresponds to a period of $\sim 2$ 
billion years. This might be considered the minimum time a backsplash 
halo spends within the virial radius of the host halo. We can compare 
this to the time scale $t_{\rm dyn}$ for a subhalo to complete one
circular orbit at the virial radius;

\begin{equation}
  t_{\rm dyn}=\sqrt{\frac{3\pi}{G\rho}}\approx\sqrt{\frac{3\pi}{G\Delta_{\rm vir}}}\rho_{\rm b}^{-1/2},\label{equ001:DT_SatAgeDistance}
\end{equation} 

\noindent where we used $\rho\approx\Delta_{\rm vir}\rho_{\rm b}$.
This leads to $t_{\rm dyn}\approx 6\times 10^9$yr. Therefore, the
minimum time $t_{\rm inside}$ a subhalo spends inside its host is
approximately $1/3^{\rm rd}$ the time it would take to complete one
orbit at the virial radius. This suggests that backsplash haloes are
on preferentially radial orbits, and explains why subhalos accreted at
earlier times are preferably found outside $R_{\rm vir}$
today.\\

We conclude that an age-distance relation akin to the one reported by
\citet{2004MNRAS.355..819G} is valid only for ``recently'' accreted
subhalos. According to \Fig{fig:ZredDist_backnbound}, there is a
clear correlation between a subhalo's age and its distance apparent 
in CDM2/WDM2 and CDM3/WDM3, the two youngest sets of hosts in our
sample, and this correlation is apparent for objects accreted after 
$z=0.2$. However, as we note above, there is some evidence that there
may be distinct subhalo populations, separated according to their
accretion redshift, that follow distinct relations and inverse relations.

It is worth noting that \citet{2004MNRAS.355..819G} observed an
age-distance relation over a much greater time span ranging from
$z\sim0.9$ down to $z\sim0.3-0.4$. These authors identify all subhalos
at $z$=1 and track them forward in time, which is equivalent to our
approach (we start tracking subhalos at the formation redshift of the
host, which varies between $z \simeq 0.8$ for CDM1/WDM1 to $z\simeq
0.2$ for CDM3/WDM3). We note that accretion in their data appears to
be complete by $z\sim0.3$ (see upper panel of their Figure 15).
However, it remains unclear why there is no further accretion apparent
in that plot for smaller redshifts as the fraction of accreted
subhalos increases at least down to $z\sim 0.1$ (cf. upper left panel
in their Figure 12). \citet{2004ogci.conf..513M} deduced from their
analysis that any age-distance relation present in their data had to
be very weak with a large scatter. We note that these authors identify
subhalos at $z$=0 and track their merger trees backwards in time.

It is clear from our analysis that any conclusions we draw must be
tentative -- if we are to gain greater insight into the age-distance
relation then we must draw upon a larger (i.e. statistical) sample of
host halos. Nevertheless we note that any age-distance relation is
apparent in the subhalo populations in both the CDM and WDM models.\\

It is notable that there are no counterparts in the WDM1 system to the
``old'' backsplash population we observe in the CDM1 model.  To better
understand why this might so, we checked the distribution of infall
velocities for subhalos \emph{that were accreted at early times} in
each of host (i.e. $z>0.75$ for CDM1 and $z>0.80$ for WDM1). We found
that the typical subhalo velocity in the CDM1 run was approximately
30\% larger than in the WDM1 run. This explains why we do not find an
``old'' backsplash population in the WDM1 run -- the typical infall
velocity of a subhalo is too low to allow it to escape the host and
become a backsplash halo. It is interesting to note that this
behaviour is peculiar to the CDM1/WDM1 set of hosts, but we consider
it statistically insignificant.

\subsection{Relative Velocities}
\label{sec:relative_velocities}
\begin{figure}
  \includegraphics[width=84mm]{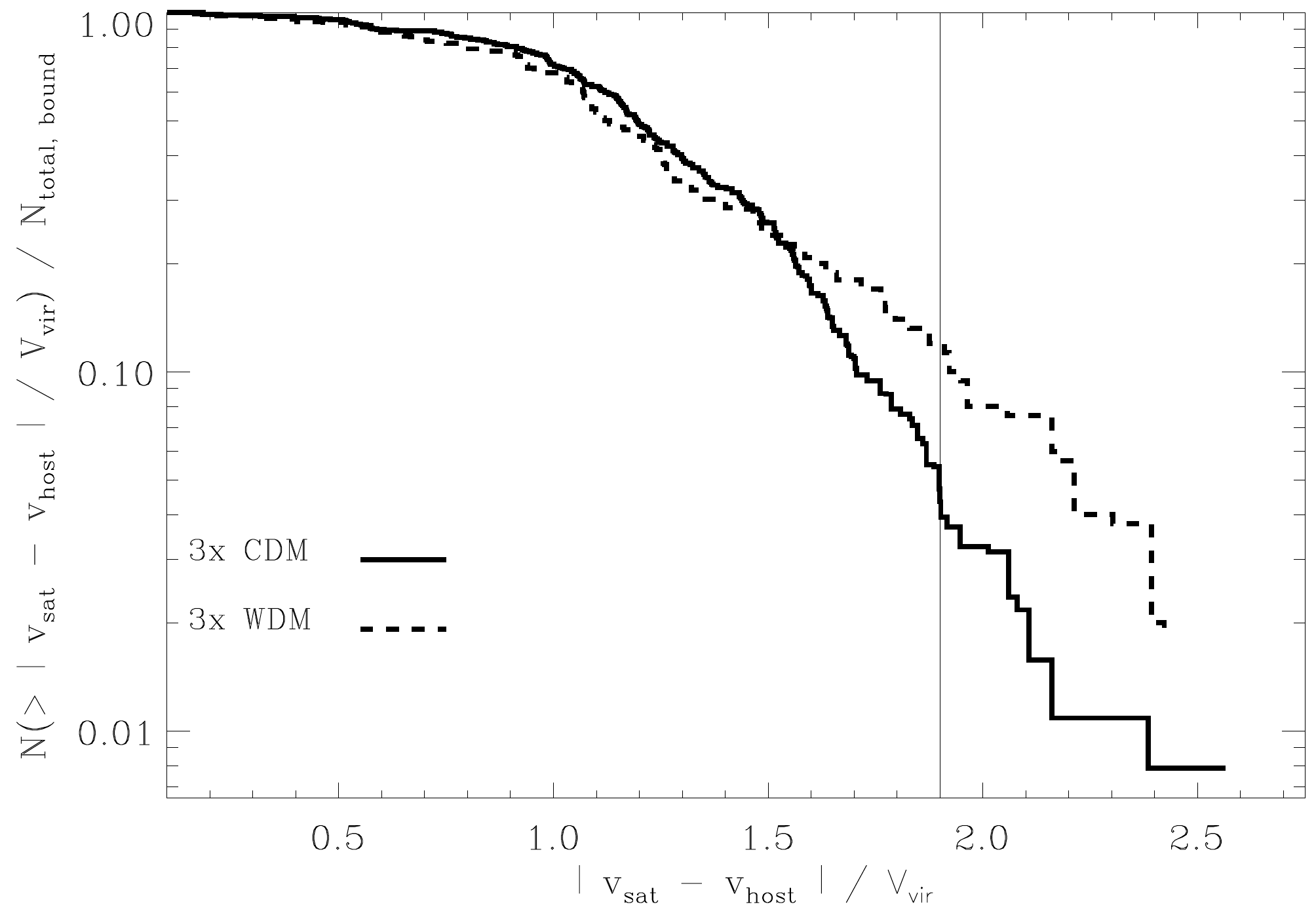}
  \caption{Cumulative distribution of relative velocity between all
    interior subhalos and their respective host. The thin vertical line
    is representative of the collisional speed of the ``Bullet''
    cluster.}
  \label{fig:CumVrel_bound_logX}
\end{figure}

\begin{figure}
  \includegraphics[width=84mm]{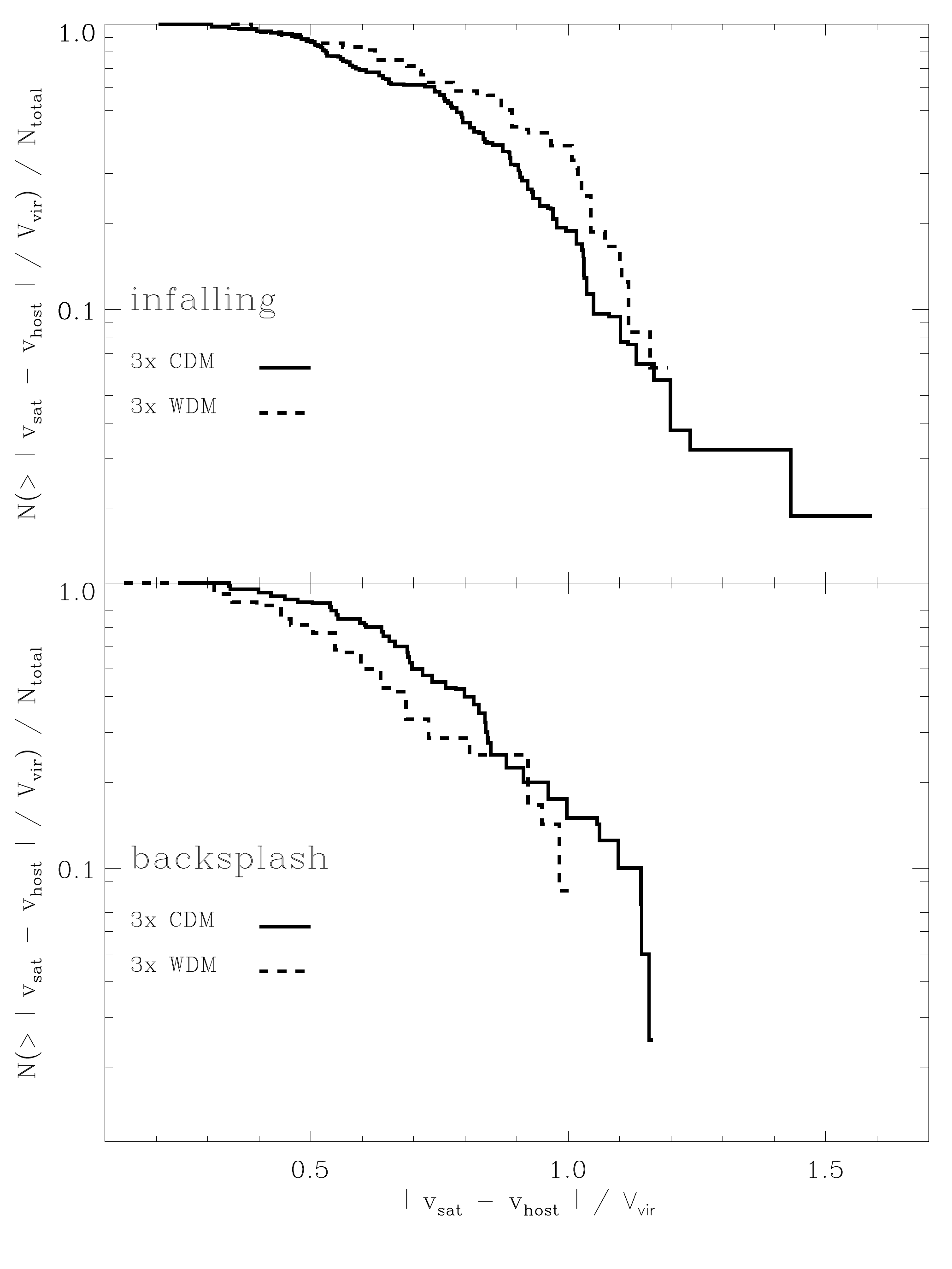}
  \caption{Same as \Fig{fig:CumVrel_bound_logX} but for the backsplash and 
	infalling population, respectively.}
  \label{fig:CumVrel_infback_logX}
\end{figure}

\begin{figure}
  \includegraphics[width=84mm]{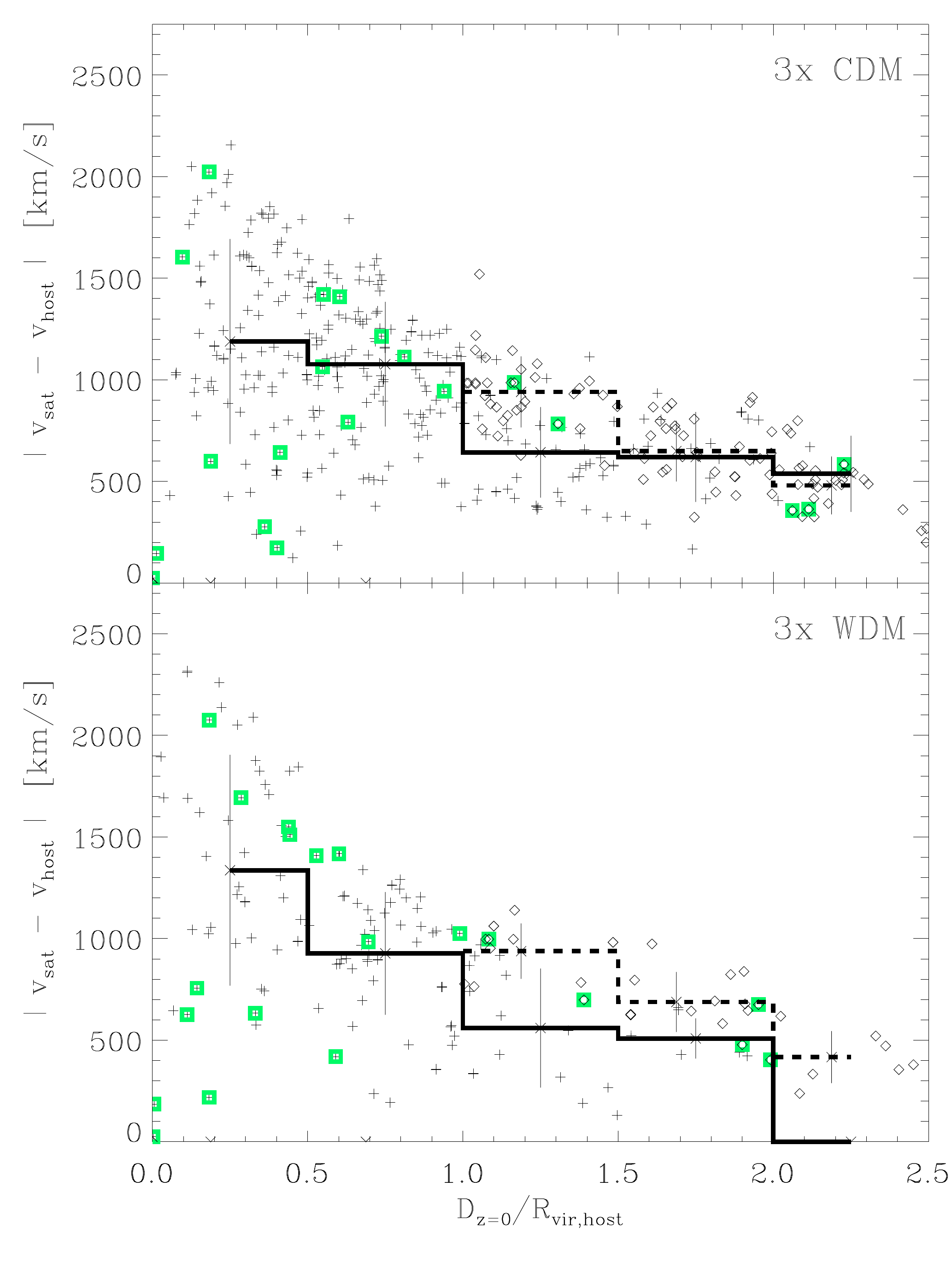}
  \caption{The absolute value of the relative velocities between the
    (sub)halos and their respective host versus their distance at the
    present time is shown. Diamonds represent the infalling population
    and black crosses denote the interior and backsplash populations. The
    thick solid lines are the mean values with error bars showing the
    standard deviation in each bin. For clarity, these are offset for
    the infalling halos. The squares denote the 20 most massive
    subhalos.}
  \label{fig:VrelDist}
\end{figure}

So far we have compared and contrasted the spatial distribution and
subhalos in the CDM and WDM models, and the degree to which these
subhalos suffer mass loss. It is also of interest to ask whether the
kinematics of subhalos in the respective models differs. We already
learnt that the early infalling population in CDM1 is marginally
faster than their counterpart in WDM1. Hence, if there are systematic
differences between the models, what might the implication be for a
system such as the ``Bullet Cluster''?  This is an extremely high
velocity merger between two galaxy clusters
\citep[][]{2007ApJ...661L.131M, 2002ApJ...567L..27M} and it has
prompted discussion as to whether such high relative velocities (of
order $\sim$ 4500 km/sec) can be accommodated within the CDM model
\citep{2007arXiv0709.3572N, 2007arXiv0704.0381A, 2007MNRAS.380..911S,
  2006MNRAS.370L..38H, 2005MNRAS.356.1327G}. We might expect such high
relative velocities to be sensitive to the large scale gravitational
field.\footnote{We caution the reader that the ``Bullet'' cluster is a
  system where the host (and the ``Bullet'' itself) is an order of
  magnitude more massive than the hosts (and satellites) presented in
  this study. While the existence of the ``Bullet'' cluster may serve
  as a motivation for the study of relative velocities any conclusions
  drawn from our results are to be extrapolated to the ``Bullet''
  system with care. This is especially so because the differences between 
  CDM and WDM are less prominent on scales corresponding to the ``Bullet'' 
  cluster.} Furthermore, if there are differences in the relative velocity
distributions in the WDM and CDM models, this could also allow limits to be 
placed on the nature of the dark matter.

In \Fig{fig:CumVrel_bound_logX} we plot the cumulative distribution of
relative velocities $(V_{\rm sat} - V_{\rm host})$ for all interior
subhalos, where $V_{\rm sat}$ and $V_{\rm host}$ are the
centre-of-mass velocities of all particles inside the virial radius of
the subhalo and the host, respectively. Relative velocities have
been normalised to the circular velocity $V_{\rm vir}$ of the host at
the virial radius. If we compute this quantity for the ``Bullet
Cluster'' using the estimate of the mass deduced from weak lensing
\citep{2004ApJ...604..596C}, the normalised collision speed $(V_{\rm
  sat}-V_{\rm host})/V_{\rm vir}$ is approximately $1.9$ (and shown as
a thin vertical line). This figure reveals that $\sim$6\% of subhalos
in the CDM model have normalised relative velocities in excess of
$1.9$, compared to $\sim$10\% in the WDM model. In other words, the
probability of a high-speed encounter is greater in the WDM model than
in the CDM model.

We have computed the same distributions for infalling and backsplash
(sub)halos and the results are shown in
\Fig{fig:CumVrel_infback_logX}.  There we observe that the infalling
population in the WDM model is marginally faster than its CDM
counterpart. However, we also note that the fastest infalling
satellitecan always be found in the CDM model. Or in other words, in
WDM there are more infalling subhalos with relative velocities up to
about $1.2\times V_{\rm vir}$, but in CDM there exists the odd
satellite with a velocity as high as $\sim 1.5\times V_{\rm vir}$. One
potential explanation of this may be that subhalos suffer ``dynamical
friction within the filaments''. As shown by
\citet{2003MNRAS.345.1285K}, more mass in CDM filaments is found in
gravitationally bound objects whereas the mass in a WDM filament is
more uniformly distributed, which may lead to enhanced dynamical
friction. Therefore, subhalos falling along filaments may have their
infall velocities reduced and hence we a) do not find exceedingly fast
subhalos and b) observe an increase in the number of objects in the
range 0.75-1.2 $V_{\rm vir}$. The situation though is different for
the backsplash population that appears to be slower than its CDM
counterpart.  

In \Fig{fig:VrelDist} we plot the (unnormalised) relative velocities of 
subhalos as a function of their present-day halocentric radii. Interior
and backsplash populations are represented by crosses while infalling 
subhalos are represented by diamonds; we also highlight the 20 most 
massive subhalos by green squares. 

There are a number of points to note in this figure. The first is that
the subhalos with the highest relative velocities are concentrated
towards the centre of the host, nestled in its potential well. The
second is that the most massive subhalos (green squares) are not
responsible for the high-velocity tail that we observe in
\Fig{fig:CumVrel_bound_logX}. The third point is that the WDM
backsplash population a) does not extend spatially as far out as for its CDM
counterpart and b) has lower velocities leading to the observed
steeper decline of the velocities with increasing clustercentric
distance. The fourth and final point is that infalling subhalos are a distinct
population kinematically, tending to have higher velocities than
backsplash galaxies \citep[see also][]{2005MNRAS.356.1327G}. We conclude
that the kinematics of subhalos is unlikely to allow us to differentiate
between the WDM and CDM models.

\section{Discussion \& Conclusions}
\label{sec:conclusions}

We have compared and contrasted the properties of subhalos orbiting in
a set of simulated galaxy cluster hosts in the CDM and WDM models. The
mass and force resolution of our simulations were sufficient to our
host halos with $\sim 10^6$ particles within the virial radius at
$z$=0, and we could follow the orbital evolution of hundreds of
subhalos in detail using outputs finely spaced in time ($\Delta t
\approx 170$~Myrs) from the formation time of the host to the present day.

Our study has revealed that many of the properties of subhalos in the
CDM model and the WDM model we have studied are similar. Subhalos in
both the CDM and WDM models are distributed anisotropically with
respect to the major axis of their host, and the phenomenon of
``backsplash'' halos is common to both models. Other studies have
shown that low-mass halos in WDM models tend to be less centrally
concentrated than their counterparts in the CDM model
\citep{2007arXiv0709.4027C}, and this leads to enhanced mass loss via
tidal stripping for subhalos in WDM models. We find no evidence for a
well pronounced correlation between the age of a subhalo and its
present day halocentric radius in either model. Interestingly, we find
that subhalos in the WDM model are likely to have higher (infall)
velocities than in the CDM model.\\

Our results nevertheless suggest that it is unlikely that the spatial
distribution and kinematics of subhalos can be used to differentiate
between the CDM and WDM models at $z$=0. It might be possible to
detect differences at higher redshifts, when the effect of the
filtering mass is more pronounced
\citep[e.g. ][]{Power.etal.2008}. Furthermore, it might be possible to
detect differences in the stellar populations and star formation
histories of the satellite galaxies that are hosted by subhalos
\citep[e.g.][]{2007Sci...317.1527G}, which will be sensitive to the
mass assembly histories of the (sub)halos. However, such measures
depend explicitly on the veracity of galaxy formation modelling, and
so it seems more likely that estimates of the small scale power
spectrum deduced from the Lyman-$\alpha$ forest
\citep[e.g.][]{2007arXiv0709.0131V} may provide stronger
constraints. Nevertheless, it is important to consider the various
strands of observational evidence when piecing together the dark
matter puzzle.

\section*{Acknowledgements} 
\label{sec:acknowledgements} 

AK acknowledges funding through the Emmy Noether Programme by the DFG
(KN 755/1). CP and BKG acknowledge the support of the Australian Research
Council supported ``Commonwealth Cosmology
Initiative''\footnote{\texttt{http://www.thecci.org}}, grant DP
0665574. The simulations were carried out on the Swinburne
Supercomputer at the Centre for Astrophysics~\& Supercomputing,
Swinburne University. All of the analyses was carried out on the
Sanssouci cluster at the AIP.

\vspace{1cm} \bsp

\bibliographystyle{mn2e}
\bibliography{paper}

\begin{thebibliography}{}

\bibitem[\protect\citeauthoryear{{Abazajian}}{{Abazajian}}{2006}]{2006PhRvD..7%
3f3513A}
{Abazajian} K.,  2006, \prd, 73, 063513

\bibitem[\protect\citeauthoryear{{Agustsson} \& {Brainerd}}{{Agustsson} \&
  {Brainerd}}{2006}]{2006ApJ...650..550A}
{Agustsson} I.,  {Brainerd} T.~G.,  2006, \apj, 650, 550

\bibitem[\protect\citeauthoryear{{Agustsson} \& {Brainerd}}{{Agustsson} \&
  {Brainerd}}{2007}]{2007arXiv0704.3441A}
{Agustsson} I.,  {Brainerd} T.~G.,  2007, ArXiv e-prints, 704

\bibitem[\protect\citeauthoryear{{Angus} \& {McGaugh}}{{Angus} \&
  {McGaugh}}{2007}]{2007arXiv0704.0381A}
{Angus} G.~W.,  {McGaugh} S.~S.,  2007, ArXiv e-prints, 704

\bibitem[\protect\citeauthoryear{{Avila-Reese}, {Col{\'{\i}}n}, {Valenzuela},
  {D'Onghia} \& {Firmani}}{{Avila-Reese} et~al.}{2001}]{2001ApJ...559..516A}
{Avila-Reese} V.,  {Col{\'{\i}}n} P.,  {Valenzuela} O.,  {D'Onghia} E.,
  {Firmani} C.,  2001, \apj, 559, 516

\bibitem[\protect\citeauthoryear{{Bailin}, {Power}, {Norberg}, {Zaritsky} \&
  {Gibson}}{{Bailin} et~al.}{2007}]{2007arXiv0706.1350B}
{Bailin} J.,  {Power} C.,  {Norberg} P.,  {Zaritsky} D.,    {Gibson} B.~K.,
  2007, ArXiv e-prints, 706

\bibitem[\protect\citeauthoryear{{Balogh}, {Navarro} \& {Morris}}{{Balogh}
  et~al.}{2000}]{2000ApJ...540..113B}
{Balogh} M.~L.,  {Navarro} J.~F.,    {Morris} S.~L.,  2000, \apj, 540, 113

\bibitem[\protect\citeauthoryear{{Bardeen}, {Bond}, {Kaiser} \&
  {Szalay}}{{Bardeen} et~al.}{1986}]{1986ApJ...304...15B}
{Bardeen} J.~M.,  {Bond} J.~R.,  {Kaiser} N.,    {Szalay} A.~S.,  1986, \apj,
  304, 15

\bibitem[\protect\citeauthoryear{{Bento}, {Bertolami}, {Rosenfeld} \&
  {Teodoro}}{{Bento} et~al.}{2000}]{2000PhRvD..62d1302B}
{Bento} M.~C.,  {Bertolami} O.,  {Rosenfeld} R.,    {Teodoro} L.,  2000, \prd,
  62, 041302

\bibitem[\protect\citeauthoryear{{Bland-Hawthorn} \&
  {Peebles}}{{Bland-Hawthorn} \& {Peebles}}{2006}]{2006Sci...313..311B}
{Bland-Hawthorn} J.,  {Peebles} P.~J.~E.,  2006, Science, 313, 311

\bibitem[\protect\citeauthoryear{{Bode}, {Ostriker} \& {Turok}}{{Bode}
  et~al.}{2001}]{2001ApJ...556...93B}
{Bode} P.,  {Ostriker} J.~P.,    {Turok} N.,  2001, \apj, 556, 93

\bibitem[\protect\citeauthoryear{{Bullock}}{{Bullock}}{2001}]{2001astro.ph.110%
05B}
{Bullock} J.~S.,  2001, ArXiv Astrophysics e-prints (0111005)

\bibitem[\protect\citeauthoryear{{Clowe}, {Gonzalez} \& {Markevitch}}{{Clowe}
  et~al.}{2004}]{2004ApJ...604..596C}
{Clowe} D.,  {Gonzalez} A.,    {Markevitch} M.,  2004, \apj, 604, 596

\bibitem[\protect\citeauthoryear{{Colin}, {Valenzuela} \&
  {Avila-Reese}}{{Colin} et~al.}{2007}]{2007arXiv0709.4027C}
{Colin} P.,  {Valenzuela} O.,    {Avila-Reese} V.,  2007, ArXiv e-prints, 709

\bibitem[\protect\citeauthoryear{{Dalcanton} \& {Hogan}}{{Dalcanton} \&
  {Hogan}}{2001}]{2001ApJ...561...35D}
{Dalcanton} J.~J.,  {Hogan} C.~J.,  2001, \apj, 561, 35

\bibitem[\protect\citeauthoryear{{De Lucia}, {Kauffmann}, {Springel}, {White},
  {Lanzoni}, {Stoehr}, {Tormen} \& {Yoshida}}{{De Lucia}
  et~al.}{2004}]{2004MNRAS.348..333D}
{De Lucia} G.,  {Kauffmann} G.,  {Springel} V.,  {White} S.~D.~M.,  {Lanzoni}
  B.,  {Stoehr} F.,  {Tormen} G.,    {Yoshida} N.,  2004, \mnras, 348, 333

\bibitem[\protect\citeauthoryear{{Diemand}, {Kuhlen} \& {Madau}}{{Diemand}
  et~al.}{2007}]{2007ApJ...667..859D}
{Diemand} J.,  {Kuhlen} M.,    {Madau} P.,  2007, \apj, 667, 859

\bibitem[\protect\citeauthoryear{{Diemand}, {Moore} \& {Stadel}}{{Diemand}
  et~al.}{2005}]{2005Natur.433..389D}
{Diemand} J.,  {Moore} B.,    {Stadel} J.,  2005, \nat, 433, 389

\bibitem[\protect\citeauthoryear{{Faltenbacher}, {Li}, {Mao}, {van den Bosch},
  {Yang}, {Jing}, {Pasquali} \& {Mo}}{{Faltenbacher}
  et~al.}{2007}]{2007ApJ...662L..71F}
{Faltenbacher} A.,  {Li} C.,  {Mao} S.,  {van den Bosch} F.~C.,  {Yang} X.,
  {Jing} Y.~P.,  {Pasquali} A.,    {Mo} H.~J.,  2007, \apjl, 662, L71

\bibitem[\protect\citeauthoryear{{Gao} \& {Theuns}}{{Gao} \&
  {Theuns}}{2007}]{2007Sci...317.1527G}
{Gao} L.,  {Theuns} T.,  2007, Science, 317, 1527

\bibitem[\protect\citeauthoryear{{Gao}, {White}, {Jenkins}, {Stoehr} \&
  {Springel}}{{Gao} et~al.}{2004}]{2004MNRAS.355..819G}
{Gao} L.,  {White} S.~D.~M.,  {Jenkins} A.,  {Stoehr} F.,    {Springel} V.,
  2004, \mnras, 355, 819

\bibitem[\protect\citeauthoryear{{Gentile}, {Tonini} \& {Salucci}}{{Gentile}
  et~al.}{2007}]{2007AaA...467..925G}
{Gentile} G.,  {Tonini} C.,    {Salucci} P.,  2007, \aap, 467, 925

\bibitem[\protect\citeauthoryear{{Gill}, {Knebe} \& {Gibson}}{{Gill}
  et~al.}{2004}]{2004MNRAS.351..399G}
{Gill} S.~P.~D.,  {Knebe} A.,    {Gibson} B.~K.,  2004, \mnras, 351, 399

\bibitem[\protect\citeauthoryear{{Gill}, {Knebe} \& {Gibson}}{{Gill}
  et~al.}{2005}]{2005MNRAS.356.1327G}
{Gill} S.~P.~D.,  {Knebe} A.,    {Gibson} B.~K.,  2005, \mnras, 356, 1327

\bibitem[\protect\citeauthoryear{{Gill}, {Knebe}, {Gibson} \& {Dopita}}{{Gill}
  et~al.}{2004}]{2004MNRAS.351..410G}
{Gill} S.~P.~D.,  {Knebe} A.,  {Gibson} B.~K.,    {Dopita} M.~A.,  2004,
  \mnras, 351, 410

\bibitem[\protect\citeauthoryear{{Goerdt}, {Moore}, {Read}, {Stadel} \&
  {Zemp}}{{Goerdt} et~al.}{2006}]{2006MNRAS.368.1073G}
{Goerdt} T.,  {Moore} B.,  {Read} J.~I.,  {Stadel} J.,    {Zemp} M.,  2006,
  \mnras, 368, 1073

\bibitem[\protect\citeauthoryear{{Hayashi} \& {White}}{{Hayashi} \&
  {White}}{2006}]{2006MNRAS.370L..38H}
{Hayashi} E.,  {White} S.~D.~M.,  2006, \mnras, 370, L38

\bibitem[\protect\citeauthoryear{{Kang}, {van den Bosch}, {Yang}, {Mao}, {Mo},
  {Li} \& {Jing}}{{Kang} et~al.}{2007}]{2007MNRAS.378.1531K}
{Kang} X.,  {van den Bosch} F.~C.,  {Yang} X.,  {Mao} S.,  {Mo} H.~J.,  {Li}
  C.,    {Jing} Y.~P.,  2007, \mnras, 378, 1531

\bibitem[\protect\citeauthoryear{{Klypin}, {Kravtsov}, {Valenzuela} \&
  {Prada}}{{Klypin} et~al.}{1999}]{1999ApJ...522...82K}
{Klypin} A.,  {Kravtsov} A.~V.,  {Valenzuela} O.,    {Prada} F.,  1999, \apj,
  522, 82

\bibitem[\protect\citeauthoryear{{Knebe}, {Devriendt}, {Gibson} \&
  {Silk}}{{Knebe} et~al.}{2003}]{2003MNRAS.345.1285K}
{Knebe} A.,  {Devriendt} J.~E.~G.,  {Gibson} B.~K.,    {Silk} J.,  2003,
  \mnras, 345, 1285

\bibitem[\protect\citeauthoryear{{Knebe}, {Devriendt}, {Mahmood} \&
  {Silk}}{{Knebe} et~al.}{2002}]{2002MNRAS.329..813K}
{Knebe} A.,  {Devriendt} J.~E.~G.,  {Mahmood} A.,    {Silk} J.,  2002, \mnras,
  329, 813

\bibitem[\protect\citeauthoryear{{Knebe}, {Gill}, {Gibson}, {Lewis}, {Ibata} \&
  {Dopita}}{{Knebe} et~al.}{2004}]{2004ApJ...603....7K}
{Knebe} A.,  {Gill} S.~P.~D.,  {Gibson} B.~K.,  {Lewis} G.~F.,  {Ibata} R.~A.,
    {Dopita} M.~A.,  2004, \apj, 603, 7

\bibitem[\protect\citeauthoryear{{Knebe}, {Green} \& {Binney}}{{Knebe}
  et~al.}{2001}]{2001MNRAS.325..845K}
{Knebe} A.,  {Green} A.,    {Binney} J.,  2001, \mnras, 325, 845

\bibitem[\protect\citeauthoryear{{Knebe}, {Power}, {Gill} \& {Gibson}}{{Knebe}
  et~al.}{2006}]{2006MNRAS.368..741K}
{Knebe} A.,  {Power} C.,  {Gill} S.~P.~D.,    {Gibson} B.~K.,  2006, \mnras,
  368, 741

\bibitem[\protect\citeauthoryear{{Kravtsov}, {Gnedin} \& {Klypin}}{{Kravtsov}
  et~al.}{2004}]{2004ApJ...609..482K}
{Kravtsov} A.~V.,  {Gnedin} O.~Y.,    {Klypin} A.~A.,  2004, \apj, 609, 482

\bibitem[\protect\citeauthoryear{{Lacey} \& {Cole}}{{Lacey} \&
  {Cole}}{1993}]{1993MNRAS.262..627L}
{Lacey} C.,  {Cole} S.,  1993, \mnras, 262, 627

\bibitem[\protect\citeauthoryear{{Libeskind}, {Cole}, {Frenk}, {Okamoto} \&
  {Jenkins}}{{Libeskind} et~al.}{2007}]{2007MNRAS.374...16L}
{Libeskind} N.~I.,  {Cole} S.,  {Frenk} C.~S.,  {Okamoto} T.,    {Jenkins} A.,
  2007, \mnras, 374, 16

\bibitem[\protect\citeauthoryear{{Libeskind}, {Frenk}, {Cole}, {Helly},
  {Jenkins}, {Navarro} \& {Power}}{{Libeskind}
  et~al.}{2005}]{2005MNRAS.363..146L}
{Libeskind} N.~I.,  {Frenk} C.~S.,  {Cole} S.,  {Helly} J.~C.,  {Jenkins} A.,
  {Navarro} J.~F.,    {Power} C.,  2005, \mnras, 363, 146

\bibitem[\protect\citeauthoryear{{Little}, {Knebe} \& {Islam}}{{Little}
  et~al.}{2003}]{2003MNRAS.341..617L}
{Little} B.,  {Knebe} A.,    {Islam} R.~R.,  2003, \mnras, 341, 617

\bibitem[\protect\citeauthoryear{{Mamon}, {Sanchis}, {Salvador-Sol{\'e}} \&
  {Solanes}}{{Mamon} et~al.}{2004}]{2004AaA...414..445M}
{Mamon} G.~A.,  {Sanchis} T.,  {Salvador-Sol{\'e}} E.,    {Solanes} J.~M.,
  2004, \aap, 414, 445

\bibitem[\protect\citeauthoryear{{Markevitch}, {Gonzalez}, {David},
  {Vikhlinin}, {Murray}, {Forman}, {Jones} \& {Tucker}}{{Markevitch}
  et~al.}{2002}]{2002ApJ...567L..27M}
{Markevitch} M.,  {Gonzalez} A.~H.,  {David} L.,  {Vikhlinin} A.,  {Murray} S.,
   {Forman} W.,  {Jones} C.,    {Tucker} W.,  2002, \apjl, 567, L27

\bibitem[\protect\citeauthoryear{{McGaugh}, {de Blok}, {Schombert}, {Kuzio de
  Naray} \& {Kim}}{{McGaugh} et~al.}{2007}]{2007ApJ...659..149M}
{McGaugh} S.~S.,  {de Blok} W.~J.~G.,  {Schombert} J.~M.,  {Kuzio de Naray} R.,
     {Kim} J.~H.,  2007, \apj, 659, 149

\bibitem[\protect\citeauthoryear{{Milosavljevi{\'c}}, {Koda}, {Nagai}, {Nakar}
  \& {Shapiro}}{{Milosavljevi{\'c}} et~al.}{2007}]{2007ApJ...661L.131M}
{Milosavljevi{\'c}} M.,  {Koda} J.,  {Nagai} D.,  {Nakar} E.,    {Shapiro}
  P.~R.,  2007, \apjl, 661, L131

\bibitem[\protect\citeauthoryear{{Moore}, {Diemand} \& {Stadel}}{{Moore}
  et~al.}{2004}]{2004ogci.conf..513M}
{Moore} B.,  {Diemand} J.,    {Stadel} J.,  2004, in {Diaferio} A.,  ed., IAU
  Colloq. 195: Outskirts of Galaxy Clusters: Intense Life in the Suburbs {On
  the age-radius relation and orbital history of cluster galaxies}.
pp 513--518

\bibitem[\protect\citeauthoryear{{Moore}, {Ghigna}, {Governato}, {Lake},
  {Quinn}, {Stadel} \& {Tozzi}}{{Moore} et~al.}{1999}]{1999ApJ...524L..19M}
{Moore} B.,  {Ghigna} S.,  {Governato} F.,  {Lake} G.,  {Quinn} T.,  {Stadel}
  J.,    {Tozzi} P.,  1999, \apjl, 524, L19

\bibitem[\protect\citeauthoryear{{Navarro}, {Hayashi}, {Power}, {Jenkins},
  {Frenk}, {White}, {Springel}, {Stadel} \& {Quinn}}{{Navarro}
  et~al.}{2004}]{2004MNRAS.349.1039N}
{Navarro} J.~F.,  {Hayashi} E.,  {Power} C.,  {Jenkins} A. .~R.,  {Frenk}
  C.~S.,  {White} S.~D.~M.,  {Springel} V.,  {Stadel} J.,    {Quinn} T.~R.,
  2004, \mnras, 349, 1039

\bibitem[\protect\citeauthoryear{{Nusser}}{{Nusser}}{2007}]{2007arXiv0709.3572%
N}
{Nusser} A.,  2007, ArXiv e-prints, 709

\bibitem[\protect\citeauthoryear{{Power}, {Bland-Hawthorn} \& {Lewis}}{{Power}
  et~al.}{2008}]{Power.etal.2008}
{Power} C.,  {Bland-Hawthorn} J.,    {Lewis} G.,  2008, in preparation, 0, 0

\bibitem[\protect\citeauthoryear{{Reed}, {Governato}, {Quinn}, {Gardner},
  {Stadel} \& {Lake}}{{Reed} et~al.}{2005}]{2005MNRAS.359.1537R}
{Reed} D.,  {Governato} F.,  {Quinn} T.,  {Gardner} J.,  {Stadel} J.,    {Lake}
  G.,  2005, \mnras, 359, 1537

\bibitem[\protect\citeauthoryear{{Reed}, {Governato}, {Verde}, {Gardner},
  {Quinn}, {Stadel}, {Merritt} \& {Lake}}{{Reed}
  et~al.}{2005}]{2005MNRAS.357...82R}
{Reed} D.,  {Governato} F.,  {Verde} L.,  {Gardner} J.,  {Quinn} T.,  {Stadel}
  J.,  {Merritt} D.,    {Lake} G.,  2005, \mnras, 357, 82

\bibitem[\protect\citeauthoryear{{Sales}, {Navarro}, {Abadi} \&
  {Steinmetz}}{{Sales} et~al.}{2007}]{2007MNRAS.379.1464S}
{Sales} L.~V.,  {Navarro} J.~F.,  {Abadi} M.~G.,    {Steinmetz} M.,  2007,
  \mnras, 379, 1464

\bibitem[\protect\citeauthoryear{{Seljak} \& {Zaldarriaga}}{{Seljak} \&
  {Zaldarriaga}}{1996}]{1996ApJ...469..437S}
{Seljak} U.,  {Zaldarriaga} M.,  1996, \apj, 469, 437

\bibitem[\protect\citeauthoryear{{Shaw}, {Weller}, {Ostriker} \& {Bode}}{{Shaw}
  et~al.}{2007}]{2007ApJ...659.1082S}
{Shaw} L.~D.,  {Weller} J.,  {Ostriker} J.~P.,    {Bode} P.,  2007, \apj, 659,
  1082

\bibitem[\protect\citeauthoryear{{Spergel} \& {Steinhardt}}{{Spergel} \&
  {Steinhardt}}{2000}]{2000PhRvL..84.3760S}
{Spergel} D.~N.,  {Steinhardt} P.~J.,  2000, Physical Review Letters, 84, 3760

\bibitem[\protect\citeauthoryear{{Spergel}}{{Spergel}}{2007}]{2007ApJS..170..3%
77S}
{Spergel} D.~N. e.~a.,  2007, \apjs, 170, 377

\bibitem[\protect\citeauthoryear{{Springel} \& {Farrar}}{{Springel} \&
  {Farrar}}{2007}]{2007MNRAS.380..911S}
{Springel} V.,  {Farrar} G.~R.,  2007, \mnras, 380, 911

\bibitem[\protect\citeauthoryear{{Springel}, {White}, {Jenkins}, {Frenk},
  {Yoshida}, {Gao}, {Navarro}, {Thacker}, {Croton}, {Helly}, {Peacock}, {Cole},
  {Thomas}, {Couchman}, {Evrard}, {Colberg} \& {Pearce}}{{Springel}
  et~al.}{2005}]{2005Natur.435..629S}
{Springel} V.,  {White} S.~D.~M.,  {Jenkins} A.,  {Frenk} C.~S.,  {Yoshida} N.,
   {Gao} L.,  {Navarro} J.,  {Thacker} R.,  {Croton} D.,  {Helly} J.,
  {Peacock} J.~A.,  {Cole} S.,  {Thomas} P.,  {Couchman} H.,  {Evrard} A.,
  {Colberg} J.,    {Pearce} F.,  2005, \nat, 435, 629

\bibitem[\protect\citeauthoryear{{Tasitsiomi}, {Kravtsov}, {Gottl{\"o}ber} \&
  {Klypin}}{{Tasitsiomi} et~al.}{2004}]{2004ApJ...607..125T}
{Tasitsiomi} A.,  {Kravtsov} A.~V.,  {Gottl{\"o}ber} S.,    {Klypin} A.~A.,
  2004, \apj, 607, 125

\bibitem[\protect\citeauthoryear{{Tormen}}{{Tormen}}{1997}]{1997MNRAS.290..411%
T}
{Tormen} G.,  1997, \mnras, 290, 411

\bibitem[\protect\citeauthoryear{{Viel}, {Becker}, {Bolton}, {Haehnelt},
  {Rauch} \& {Sargent}}{{Viel} et~al.}{2007}]{2007arXiv0709.0131V}
{Viel} M.,  {Becker} G.~D.,  {Bolton} J.~S.,  {Haehnelt} M.~G.,  {Rauch} M.,
  {Sargent} W.~L.~W.,  2007, ArXiv e-prints, 709

\bibitem[\protect\citeauthoryear{{Viel}, {Lesgourgues}, {Haehnelt}, {Matarrese}
  \& {Riotto}}{{Viel} et~al.}{2006}]{2006PhRvL..97g1301V}
{Viel} M.,  {Lesgourgues} J.,  {Haehnelt} M.~G.,  {Matarrese} S.,    {Riotto}
  A.,  2006, Physical Review Letters, 97, 071301

\bibitem[\protect\citeauthoryear{{Wang} \& {White}}{{Wang} \&
  {White}}{2007}]{2007MNRAS.380...93W}
{Wang} J.,  {White} S.~D.~M.,  2007, \mnras, 380, 93

\bibitem[\protect\citeauthoryear{{Warnick} \& {Knebe}}{{Warnick} \&
  {Knebe}}{2006}]{2006MNRAS.369.1253W}
{Warnick} K.,  {Knebe} A.,  2006, \mnras, 369, 1253

\bibitem[\protect\citeauthoryear{{Warnick}, {Knebe} \& {Power}}{{Warnick}
  et~al.}{2008}]{Warnick.Knebe.Power.2007}
{Warnick} K.,  {Knebe} A.,    {Power} C.,  2008, \mnras\ submitted, 544

\bibitem[\protect\citeauthoryear{{Willman}, {Governato}, {Dalcanton}, {Reed} \&
  {Quinn}}{{Willman} et~al.}{2004}]{2004MNRAS.353..639W}
{Willman} B.,  {Governato} F.,  {Dalcanton} J.~J.,  {Reed} D.,    {Quinn} T.,
  2004, \mnras, 353, 639

\bibitem[\protect\citeauthoryear{{Zentner} \& {Bullock}}{{Zentner} \&
  {Bullock}}{2003}]{2003ApJ...598...49Z}
{Zentner} A.~R.,  {Bullock} J.~S.,  2003, \apj, 598, 49

\bibitem[\protect\citeauthoryear{{Zentner}, {Kravtsov}, {Gnedin} \&
  {Klypin}}{{Zentner} et~al.}{2005}]{2005ApJ...629..219Z}
{Zentner} A.~R.,  {Kravtsov} A.~V.,  {Gnedin} O.~Y.,    {Klypin} A.~A.,  2005,
  \apj, 629, 219

\end{thebibliography}

\end{document}